\documentclass[journal]{IEEEtran}
\usepackage{float}
\usepackage{graphicx}
\usepackage{subfigure}
\usepackage{epstopdf}
\usepackage{color}
\usepackage{amssymb}
\usepackage{amsmath}
\usepackage{algorithm}
\usepackage{algorithmic}
\usepackage{bm}
\usepackage[colorlinks,citecolor=blue,linkcolor=blue,urlcolor=blue]{hyperref}
\usepackage{stfloats}
\begin{document}
\title{Rethinking Hardware Impairments in Multi-User Systems: Can FAS Make a Difference?}
\author{Junteng Yao, Tuo Wu, Liaoshi Zhou, Ming Jin, Cunhua Pan, Maged Elkashlan, \\ Fumiyuki Adachi, \emph{Life Fellow}, \emph{IEEE},  George K. Karagiannidis,  \emph{Fellow},  \emph{IEEE}, \\  Naofal Al-Dhahir, \emph{Fellow, IEEE},   and Chau Yuen, \emph{Fellow, IEEE}

\thanks{(\textit{Corresponding author: Tuo Wu.})}

\thanks{J. Yao, L. Zhou, and M. Jin are with the faculty of Electrical Engineering and Computer Science, Ningbo University, Ningbo 315211, China (E-mail: $\rm \{yaojunteng, 2311100202, jinming\}@nbu.edu.cn$ ).}

\thanks{T. Wu and C. Yuen are with the School of Electrical and Electronic Engineering, Nanyang Technological University, 639798, Singapore (E-mail: $\rm \{tuo.wu, chau.yuen\}@ntu.edu.sg$).}
	
 \thanks{C. Pan is with the National Mobile Communications Research Laboratory, Southeast University, Nanjing 210096, China (E-mail: $\rm cpan@seu.edu.cn$)}

\thanks{M. Elkashlan is with the School of Electronic Engineering and Computer Science at Queen Mary University of London, London E1 4NS, U.K. (E-mail: $\rm maged.elkashlan@qmul.ac.uk$). }
\thanks{F. Adachi is with the International Research Institute of Disaster Science (IRIDeS), Tohoku University, Sendai, Japan (E-mail: $\rm adachi@ecei.tohoku.ac.jp$ ).}
 \thanks{ G. K. Karagiannidis is with the Department of Electrical and Computer Engineering, Aristotle University of Thessaloniki, 54124 Thessaloniki, Greece, and also with the Artificial Intelligence $\&$ Cyber Systems Research Center, Lebanese American University (LAU), Lebanon (E-mail: $\rm geokarag@auth.gr$).}

\thanks{ Naofal Al-Dhahir is with the Department of Electrical and Computer Engineering, The University of Texas at Dallas, Richardson, TX 75080 USA (E-mail: $\rm aldhahir@utdallas.edu$).
}
}
\maketitle
\begin{abstract} In this paper, we analyze the role of fluid antenna systems (FAS) in multi-user systems with hardware impairments (HIs). Specifically, we investigate a scenario where a base station (BS) equipped with multiple fluid antennas communicates with multiple users (CUs), each equipped with a single fluid antenna. Our objective is to maximize the minimum communication rate among all users by jointly optimizing the BS's transmit beamforming, the positions of its transmit fluid antennas, and the positions of the CUs' receive fluid antennas. To address this non-convex problem, we propose a block coordinate descent (BCD) algorithm integrating semidefinite relaxation (SDR), rank-one constraint relaxation (SRCR), successive convex approximation (SCA), and majorization-minimization (MM). Simulation results demonstrate that FAS significantly enhances system performance and robustness, with notable gains when both the BS and CUs are equipped with fluid antennas. Even under low transmit power conditions, deploying FAS at the BS alone yields substantial performance gains. However, the effectiveness of FAS depends on the availability of sufficient movement space, as space constraints may limit its benefits compared to fixed antenna strategies. Our findings highlight the potential of FAS to mitigate HIs and enhance multi-user system performance, while emphasizing the need for practical deployment considerations.
	
\end{abstract}
\section{Introduction}
\IEEEPARstart{O}{ver} the last 2 decades, multiple-input multiple-output (MIMO) has emerged as a powerful technology for wireless communication networks, capable of providing large diversity gains and degrees of freedom (DoFs) to improve the quality of service (QoS) and spectral efficiency  \cite{CXWang23,EGLarsson2014,ZWang2024}. However, with the increasingly stringent demands of the next-generation networks,  the performance gains of conventional MIMO systems may be limited when the channel conditions are poor, due to the static and discrete arrangement of fixed-position antennas (FPAs). To address this issue, the fluid antenna system (FAS) \cite{TWu2024,XLai24,KKWong21,JYao24,CWang24,Yao20241}, sometimes also known as the movable antenna system (MAS) \cite{LZhu241,WMa241,YGao24,LZhu20242,GHu2024,WMei2024}, has been considered as a promising technology to enhance the communication performance due to its ability of selecting the best positions of the antennas. FAS effectively has antenna position reconfigurable features, allowing it to fully exploit the channel variations in the spatial domain and boost the communication performance compared to conventional FPA systems. This is achieved by utilizing a liquid-based antenna or reconfigurable pixel-based antenna \cite{SShen2017,FJiang2022}.

The above-mentioned benefits of FAS have motivated researchers to investigate various properties of FAS-assisted wireless communication systems, particularly in the areas of channel modeling \cite{KKWong21,KKWong2022,MKhammassi23,PR24}, performance analysis \cite{CSkouroumounis2023,WKNew20241,JD2024,HXu20242,LTlebaldiyeva2023,JZheng2024},  and performance optimization \cite{New2024,JTang2024,YYe2024,ZCheng2024,GHu20242,NLi2024,JYao20242}. Wong \emph {et al} firstly introduced FAS to wireless communication systems, and proposed a spatial correlation model to analyze the outage probability of the single-user FAS \cite{KKWong21}. To effectively imitate the actual antenna in FAS, Wong \emph {et al} further proposed a constant correlation model \cite{KKWong2022}. Later in \cite{MKhammassi23},  Khammassi \emph{et al} proposed a two-stage approximation channel model with eigenvalue decomposition  to improve the performance analysis of FAS. To effectively strike a balance between the mathematical tractability and accuracy, Ramirez-Espinosa \emph{et al} presented a block-correlation model, which approximates spatial correlation by using block-diagonal matrices \cite{PR24}. 

Based on the above-mentioned models, Skouroumounis \emph{et al} analyzed the impact of the channel estimation and the port selection on the outage performance in large-scale FAS \cite{CSkouroumounis2023}. New \emph{et al} derived closed-form expressions for the outage probability and diversity gain, which enhances the understanding of the performance limit of FAS  \cite{WKNew20241}. Vega-Sánchez \emph{et al} in \cite{JD2024} investigated the secrecy outage probability of FAS by approximating the end-to-end signal-to-noise-ratio (SNR) distributions of the legitimate user and eavesdropper. Xu \emph{et al} studied the outage probability of a two-user FAS, where two users employ FAS to suppress mutual interferences\cite{HXu20242}. Non-orthogonal multiple access (NOMA) can effectively achieve interference cancellation, and integrating NOMA into FAS can significantly reduce the outage probability of FAS. Therefore, \cite{LTlebaldiyeva2023} and \cite{JZheng2024} investigated the outage probabilities of cooperative NOMA FAS and NOMA FAS with short-packet communications, respectively.

The optimization of FAS primarily focuses on the port selection or antennas' positions. New \emph{et al} studied the joint  one-dimensional (1D) port selection and power allocation scheme for orthogonal multiple access (OMA) and NOMA FAS, showing the superiority of FAS compared with conventional FPA  \cite{New2024}. Later, Wang \emph{et al} proposed a two-dimensional (2D) port selection scheme in FAS-assisted  integrated sensing and communications (ISAC) systems, where the transmitter employs FAS to improve communication and sensing performance simultaneously  \cite{CWang24}. Exploring how FAS can protect the legitimate information, Tang \emph{et al} formulated antennas' positions optimization problem in secure communication systems \cite{JTang2024}. Ye \emph{et al}  \cite{YYe2024} considered the scenario that the transmitter only can obtain the statistical channel state information (CSI), and then optimized the antennas' positions at the transmitter to improve communication rate. Furthermore, \cite{GHu20242} and \cite{NLi2024} further investigated the antennas' positions optimization problems for the downlink and uplink multi-user systems, respectively. Considering  the reconfigurability of reconfigurable intelligent surfaces (RIS), Yao \emph{et al} combined RIS with FAS to achieve performance improvement \cite{JYao20242}.

The aforementioned studies, while presenting promising and superior aspects of FAS from various perspectives, often assume ideal hardware at both the transmitters and the receivers. However, this assumption may not always hold true in practical wireless communications systems. In reality, both transmitters and receivers are inevitably subject to HIs (HIs), arising from factors such as amplifier non-linearities, amplitude/phase imbalance, phase noise, finite-resolution quantization, etc. 

Therefore, it is crucial to explore the impact of HIs on system performance. To effectively analyze wireless systems with HIs, Bjärnson \emph{et al}  in \cite{EB2014} have proven that the HIs can be modeled as an additive Gaussian distribution, whose variance is proportional to the signal power. Inspired by \cite{EB2014}, researchers have recently conducted some studies on this topic \cite{ZLiu2020,ZXing2021,GZhou2021,AP2021,JWang2023,JFang2023,JDai2024,QLi2024,ZPeng2024,HLi2024}. Liu \emph{et al} derived the outage probability of simultaneous wireless information, and power transfer (SWIPT) based two-way amplify-and-forward (AF) relay networks, and obtained the maximum tolerable level of HIs under a certain QoS constraint\cite{ZLiu2020}. Xing \emph{et al} employed RIS to suppress the interferences caused by HIs at the transceiver, and optimized the phase of the RIS to maximize the average achievable rate \cite{ZXing2021}. Zhou \emph{et al} in \cite{GZhou2021} designed the passive beamforming of RIS to maximize the secrecy rate of the RIS-assisted systems with HIs. Papazafeiropoulos \emph{et al} in \cite{AP2021} studied the impact of HIs on the cell-free massive MIMO systems, and found that the performance degradation caused by the transmitters is worse than that of the receivers. Wang \emph{et al} investigated the joint impacts of transceiver HIs, RIS phase noise and imperfect CSI on the RIS-aided MIMO systems \cite{JWang2023}. Improper Gaussian signaling (IGS)  can provide additional DoFs for the receiver, thus Fang \emph{et al} in \cite{JFang2023} proposed to use IGS to compensate the performance degradation caused by HIs. Dai \emph{et al} studied the Cramér-Rao Bound (CRB) of sensing performance minimization problem in the ISAC systems with HIs by designing the transmit beamforming of dual-function transmitter \cite{JDai2024}. Moreover, \cite{QLi2024,ZPeng2024}, and \cite{HLi2024} studied the impact of HIs on the reconfigurable holographic surfaces-assisted near-field communication systems, active RIS-assisted systems, and covert communication systems, respectively. There are a number of studies on HIs for SWIPT, AF, RIS, and other related fields, but not for FAS. To the best of our knowledge, it is the first time to exploit FAS to mitigate HIs.

\subsection{Motivation and Contributions} 
HIs  are inevitable in wireless communication systems and can significantly degrade critical performance metrics such as achievable rates and signal quality. Meanwhile,  FAS with their unique spatial diversity and adaptability, offer a promising solution for mitigating the adverse effects of HIs. This raises a fundamental question: \textbf{\textit{Can FAS make a difference in alleviating the impact of HIs?}} 

To address this intriguing yet challenging question, it is essential to first analyze how HIs affect system performance. HIs introduce additional distortion terms proportional to transmit power and channel gains, increasing the noise floor and degrading SINR, thereby limiting achievable rates. On the other hand, FAS, with its ability to dynamically adjust antenna positions, can counteract these effects by improving channel conditions and suppressing interference. This foundational understanding forms the basis for designing an effective optimization framework to enhance system performance. However, incorporating FAS into HI-impaired systems introduces significant complexity, as the added dynamics of FAS create intricate couplings between HIs and system parameters. Traditional optimization methods for addressing HIs are inadequate, necessitating novel approaches tailored to these challenges.

Based on the need for investigating how FAS can allerviate the impact of HIs in wireless systems, it is essential to consider not only single-user systems but also multi-user systems, which are more representative of practical communication networks. Multi-user systems introduce additional challenges due to inter-user interference, as users share the same spectrum resources. This makes it critical to balance the quality-of-service (QoS) among users and prevent excessively poor performance for any single user, especially in the presence of HIs. Incorporating both inter-user interference and transceiver HIs into the optimization framework further increases problem complexity, as interference terms become highly coupled with system parameters.

Additionally, while transmit fluid antennas can improve system performance by considering the channel conditions of all users simultaneously, this often leads to trade-offs that prioritize some users over others. On the other hand, receiver fluid antennas can independently optimize their positions based solely on their own channel conditions, offering better interference suppression and improved QoS. Scenarios where both transmitters and receivers are equipped with fluid antennas thus represent a more general and practical setting but also exacerbate the optimization challenges due to the increased coupling between transmitter and receiver configurations. These complexities highlight the need for advanced techniques to fully unlock the potential of FAS in multi-user systems with HIs.

Motivated by these challenges, this paper investigates the role of FAS in multi-user downlink communication systems with HIs. Specifically, we consider a system where a base station (BS) equipped with multiple fluid antennas communicates with multiple users (CUs), each equipped with a single fluid antenna. Our study focuses on maximizing the minimum communication rate among users by jointly optimizing the BS's transmit beamforming and the positions of fluid antennas at both the BS and the CUs. The main contributions of this paper are summarized as follows:

\begin{itemize}
\item  \textbf{\textit{Novel Multi-User Optimization Framework:}}
To the best of our knowledge, this is the first work to explore the impact of FAS on multi-user systems with HIs. Considering fairness among users, we formulate a max-min communication rate optimization problem subject to the BS's power constraint, the finite moving region of all fluid antennas, and the minimum separation between antennas. The joint optimization of the BS's transmit beamforming, the positions of its transmit fluid antennas, and the positions of CUs' receive fluid antennas results in a highly non-convex problem due to the intricate coupling between inter-user interference, HIs, and FAS dynamics.

\item \textbf{\textit{Advanced Optimization Algorithm:}}
To solve the highly non-convex problem, we propose a block coordinate descent (BCD)-based algorithm that decomposes the problem into three sub-problems, 
and employs several advanced techniques, i.e., semidefinite relaxation (SDR), sequential rank-one constraint relaxation (SRCR), majorization-minimization (MM), and successive convex approximation (SCA), to solve these sub-problems. Specifically,  for the BS's transmit beamforming optimization, we employ the SDR and SRCR algorithms to address the rank-one constraint of the transmit beamforming and provide superior convergence. For the transmit fluid antennas' positions optimization, we use the MM and SCA algorithms to construct the upper and lower bounds on the non-convex constraints, and transform the non-convex optimization problem into a convex form. The receive fluid antenna's position of each CU optimization problem is handled by the SCA algorithm with second-order Taylor expansion.

\item \textbf{\textit{Key Insights into the Deployment  of FAS:}}
Our study demonstrates that FAS significantly mitigates the adverse effects of HIs, achieving notable performance gains. Deploying FAS at both the BS and CUs maximizes the achievable rate and system robustness, while equipping only the BS with FAS provides substantial improvements, particularly under low transmit power conditions. However, the effectiveness of FAS is highly dependent on the availability of sufficient movement space, as constrained regions may reduce its benefits, allowing fixed antenna selection methods to outperform FAS in some cases. These insights highlight the importance of adequate movement space and strategic deployment of FAS to fully unlock its potential in mitigating HIs and enhancing multi-user system performance.
 \end{itemize}
 
\subsection{Organization} 
The rest of this paper is organized as follows. Section \ref{sec:system} presents our model of the FAS-assisted multi-user system with HIs. In Section \ref{sec:BCD}, the BCD algorithm is proposed to optimize the transmit beamforming of the BS, the transmit fluid antennas' positions at the BS, and the receive fluid antennas' positions at multiple CUs. Finally, our numerical results are provided in Section \ref{sec:Num}, and our conclusions are drawn in Section \ref{sec:conclusion}.
 
 \subsection{Notations} 
The trace, rank, Frobenius norm, 2-norm, conjugate transpose, conjugate, and transpose of the matrix $\mathbf{A}$ are denoted as $\mathrm{Tr}(\mathbf{A})$, $\mathrm{rank}(\mathbf{A})$, $\left\|\mathbf{A}\right\|_F$, $\left\|\mathbf{A}\right\|_2$, $\mathbf{A}^H$, $\mathbf{A}^{\ast}$, and $\mathbf{A}^T$, respectively; $\mathbf{A}\succeq (\succ) \mathbf{0}$ indicates that the matrix $\mathbf{A}$ is positive semidefinite (positive definite); $\Re\{x\}$ means the real part of $x$; The term $\mathbb{C}^{m\times n}$ denotes a complex matrix with size $m \times n$; $\lambda_{\max}(\mathbf{A})$ denotes the maximum eigenvalue of matrix $\mathbf{A}$. $|a|$ and $\angle a$ denote the amplitude and phase of complex scalar $a$, respectively;  $\mathbb{E}\left\{\cdot\right\}$ denotes the expectation operation; $\mathrm{diag}\{\cdot\}$ denotes the diagonalization operation; $\mathcal{CN}(\mathbf{0}, \mathbf{I})$ denotes the distribution of a circularly symmetric complex Gaussian vector with mean $\mathbf{0}$ and covariance matrix equal to the identity matrix $\mathbf{I}$.

\section{ System Model}\label{sec:system}
\begin{figure*}[t]
\centering
\includegraphics[width=4.5in]{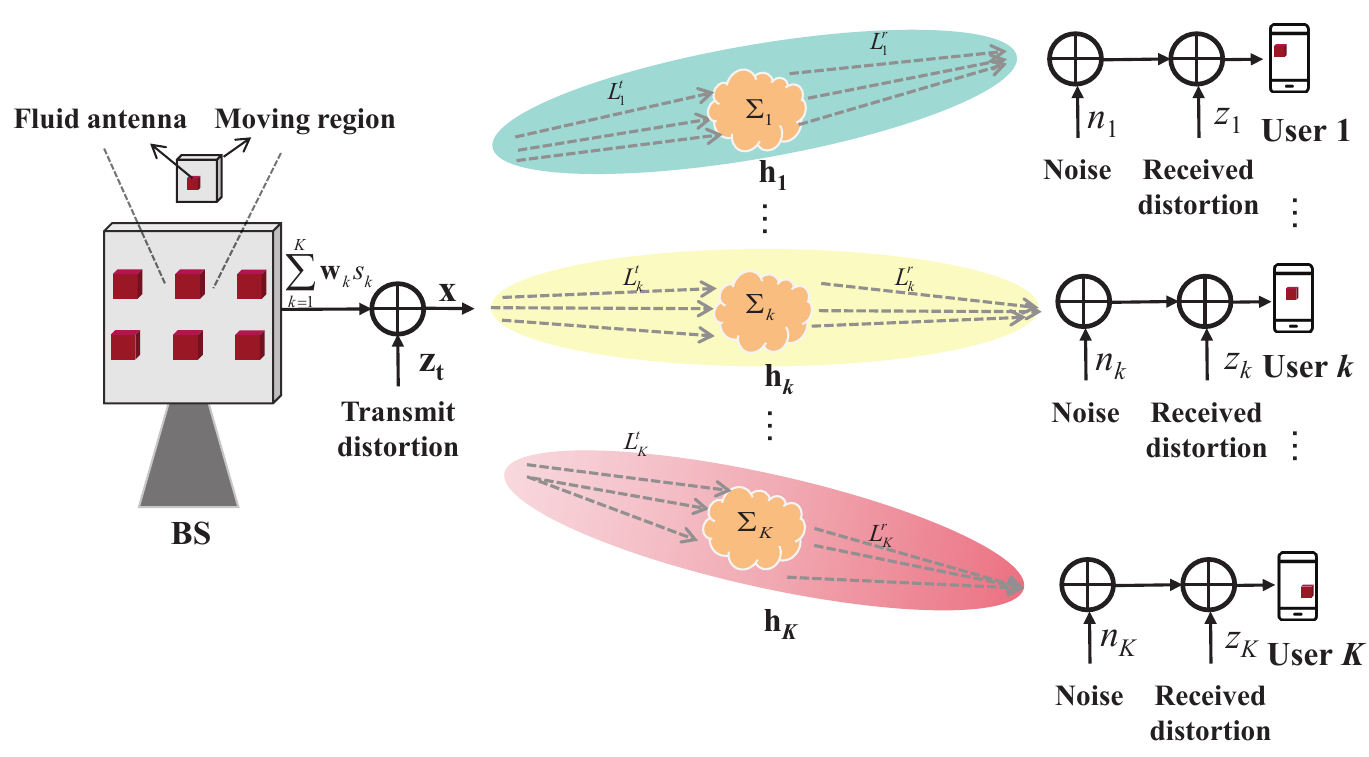}
\caption { The system model of fluid antenna-assisted multi-user communication systems.}
\label{model}
\end{figure*}
Consider a downlink FAS-assisted system where the base station (BS), equipped with $N (N\geq 2)$  fluid antennas, transmits signals to $K (K\geq 2)$ communication users (CUs), each equipped with a single fluid antenna, as shown in Fig. \ref{model}. These fluid antennas are connected to radio frequency (RF) chains via integrated waveguides or flexible cables, allowing free movement within defined ranges denoted as $\mathcal{S}_t$ for the BS and $\mathcal{S}^k_r$ for the $k$-th CU, respectively. In this paper, we use the planar far-field response model, where the angles of arrival (AoAs), angles of departure (AoDs), and path response coefficients, except for signal phases	, are the same for each channel transmission component \cite{LZhu241}. We define $\mathbf{\bar{t}} = [\mathbf{t}_1, \mathbf{t}_2, \dots, \mathbf{t}_N]$ to represent the set of positions of the transmit fluid antennas in the BS, where $\mathbf{t}_{n}=[x_n^t, y_n^t]^T,n\in\mathcal{N}=\{1,\cdots,N\}$, represents the position of the $n$-th transmit fluid antenna. Similarly, the position of the receive fluid antenna in the $k$-th CU is denoted as $\mathbf{r}_{k}=[x_k^r, y_k^r]^T,k\in \mathcal{K}=\{1,\cdots,K\}$.

Take the channel between the BS and the $k$-th CU for example, the signal propagation difference between the $n$-th transmit fluid antenna and the reference origin in the BS as $\varrho_{k,l}^t(\mathbf{t}_n) = x_{n}^{t}\sin\theta_{k,l}^t \cos\varphi_{k,l}^t + y_{n}^{t} \cos\theta_{k,l}^t$, where $\theta_{k,l}^t \in [0, \pi]$ and $\varphi_{k,l}^t \in [0, \pi]$, respectively, represent the elevation and azimuth angles of the $l$-th path ($l \in \{1, \dots, L_k^t\}$), and $L_k^t$ is the number of the transmit paths from the BS to the $k$-th CU \cite{WMa241}. Thus, the transmit field response vector (FRV) of the $n$-th transmit fluid antenna is given by \cite{LZhu241}
\begin{align}\label{eq1}
\mathbf{g}_k(\mathbf{t}_n)= \left [e^{j \frac{2\pi}{\lambda}\varrho_{k,1}^t(\mathbf{t}_n)}, \cdots, e^{j \frac{2\pi}{\lambda}\varrho_{k,L_k^t}^t(\mathbf{t}_n)}\right]^T \in\mathbb{C}^{L_k^t},
\end{align}
where $\lambda$ is the carrier wavelength. The transmit field response matrix (FRM) of the link between the BS and $k$-th CU is
\begin{align}\label{eq2}
\mathbf{G}_k\mathbf{(\bar{t})} = \left [\mathbf{g}_k(\mathbf{t}_1),\mathbf{g}_k(\mathbf{t}_2), \cdots, \mathbf{g}_k(\mathbf{t}_N)\right]\in\mathbb{C}^{L_k^t\times N}.
\end{align}
Similarly, the receive FRV in the $k$-th CU can be expressed as
\begin{align}\label{eq3}
\mathbf{f}_k(\mathbf{r}_k) = \left [e^{j \frac{2\pi}{\lambda}\varrho_{k,1}^r(\mathbf{r}_k)},\cdots,e^{j \frac{2\pi}{\lambda}\varrho_{k,L_k^r}^r(\mathbf{r}_k)}\right]^T \in\mathbb{C}^{L_k^r},
\end{align}
where $\varrho_{k,m}^r(\mathbf{r}_k) = x_k^{r}\sin\theta_{k,m}^r \cos\varphi_{k,m}^r + y_k^{r} \cos\theta_{k,m}^r$ represents the signal propagation difference of the $m$-th ($m \in \{1, \dots, L_k^r\}$) receive path between a receive fluid antenna and its referenced origin in the $k$-th CU, $L_k^r$ is the number of the receive paths from the BS to the $k$-th CU, and $\theta_{k,m}^r$ and $\varphi_{k,m}^r$ are the elevation and azimuth angles at the $k$-th CU, respectively.

Furthermore, we define the path response matrix $\bm{\Sigma}_k\in\mathbb{C}^{{L_k^r}\times {L_k^t}}$ as the responses of the transmit and receive paths between the BS and the $k$-th CU. Therefore, the channel between the BS and the $k$-th CU can be written as
\begin{align}\label{eq4}
\mathbf{h}_k=\mathbf{f}^H_k(\mathbf{r}_k)\bm{\Sigma}_k\mathbf{G}_k\mathbf{(\bar{t})}\in\mathbb{C}^{1\times N}.
\end{align}

In practical communication systems, HIs are common and can significantly impact system performance. To ensure our model accurately reflects real-world scenarios, we consider residual HIs at both the BS and the CUs. Therefore, the BS's transmit signal is formulated as
\begin{align}\label{eq5}
\mathbf{x}=\sum\limits_{k=1}^{K}\mathbf{w}_ks_k+\mathbf{z}_t,
\end{align}
where $s_k$ with $\mathbb{E}\left(|s_k|^2 \right)=1$ represents the desired signal for the $k$-th CU, $\mathbf{w}_k \in \mathbb{C}^{N \times 1}$ is the corresponding beamforming vector, respectively.
Besides, $\mathbf{z}_t \sim \mathcal{CN}\left(0,\eta\text{diag}\left\{\sum\limits_{k=1}^{K} \mathbf{w}_k\mathbf{w}_k^H\right\}\right)$ denotes the transmit distortion noise caused by HIs at the BS, with $\eta \in [0,1]$ indicating the HI coefficient \cite{GZhou2021,JWang2023}.

 Accordingly, the signal received by the $k$-th CU is given by
\begin{align}\label{eq6}
y_k=\mathbf{h}_k\mathbf{x}+n_k+z_k=\bar{y}_k+z_k,
\end{align}
where $\bar{y}_k=\mathbf{h}_k\mathbf{x}+n_k$, $n_k \sim \mathcal{CN}\left(0,\sigma^2\right)$ represents the additive white Gaussian noise (AWGN) at the $k$-th CU, $z_k \sim \mathcal{CN}\left(0,\rho_k\mathbb{E}(|\bar{y}_k|^2)\right)$ represents the received distortion noise at the $k$-th CU, and $\rho_k \in [0,1]$ is the HI coefficient at the $k$-th CU, respectively \cite{JFang2023,JDai2024}. Consequently, we have
\begin{align}\label{eq7}
\mathbb{E}(|\bar{y}_k|^2)=&\sum\limits_{i=1}^{K} \mathbf{h}_k \mathbf{w}_i \mathbf{w}_i^H \mathbf{h}_k^H\nonumber\\
&+ \eta \mathbf{h}_k \text{diag}\left\{\sum\limits_{i=1}^{K} \mathbf{w}_i \mathbf{w}_i^H\right\} \mathbf{h}_k^H
+\sigma^2,
\end{align}
Then, by expressing the second term on the right side of \eqref{eq7} as
\begin{align}\label{eq8}
\mathbf{h}_k \text{diag}\left\{\sum\limits_{i=1}^{K} \mathbf{w}_i \mathbf{w}_i^H\right\} \mathbf{h}_k^H
&=\sum\limits_{i=1}^{K} \mathbf{w}_i^H \text{diag}\{\mathbf{h}_k^H\mathbf{h}_k \} \mathbf{w}_i \nonumber \\
&=\sum\limits_{i=1}^{K} \mathrm{Tr}\left(\text{diag}\{ \mathbf{H}_k\}\mathbf{W}_i \right)
\end{align}
where $\mathbf{W}_k=\mathbf{w}_k \mathbf{w}_k^H$ and $\mathbf{H}_k=\mathbf{h}_k \mathbf{h}_k^H$, we have 
the signal-to-interference-plus-noise ratio (SINR) of the $k$-th CU written as
\begin{align}\label{eq9}
\gamma_k=\frac{\mathrm{Tr}\left(\mathbf{H}_k\mathbf{W}_k \right)}{\delta_k},
\end{align}
where
\begin{align}\label{eq10}
\delta_k = &\sum\limits_{i\neq k}^{K} \mathrm{Tr}\left(\mathbf{H}_k\mathbf{W}_i \right) +\rho_k \sum\limits_{i=1}^{K}\mathrm{Tr}\left(\mathbf{H}_k\mathbf{W}_i \right) \nonumber\\
&+\eta(1+\rho_k)\mathrm{Tr}\left(\text{diag}\{ \mathbf{H}_k\}\mathbf{W}_i \right) +(1+\rho_k)\sigma^2.
\end{align}
Hence, the achievable rate of the $k$-th CU is
\begin{align}\label{eq11}
R_k=\log_2\left(1+\gamma_k\right).
\end{align}

\section{Performance Analysis and Problem Formulation }\label{sec
}

In this section, we analyze the impact of HIs on the performance of the fluid antenna-assisted multi-user communication system and formulate an optimization problem to investigate how FAS can mitigate these effects. We begin by examining how HIs degrade system performance, then explore the mechanisms by which FAS can alleviate these issues, and finally present the optimization problem that incorporates these considerations.
\subsection{Impact of HIs on SINR} 
HIs at both the BS and the CUs introduce distortion noise, which degrades the signal quality and system performance. These impairments include amplifier nonlinearities, I/Q imbalance, and phase noise, which collectively increase the effective noise floor and interfere with the desired signals.
By expanding $\delta_k$, we can write 
\begin{align}\label{eq141}
	\delta_k &= \underbrace{ \sum_{i \neq k} \mathrm{Tr}\left( \mathbf{H}_k \mathbf{W}_i \right) }_{\text{Multi-user interference}} + \underbrace{ \rho_k \sum_{i=1}^K \mathrm{Tr}\left( \mathbf{H}_k \mathbf{W}_i \right) }_{\text{CU distortion noise}} \nonumber \\
	&\quad + \underbrace{ \eta (1 + \rho_k) \sum_{i=1}^K \mathrm{Tr}\left( \mathrm{diag}\{ \mathbf{H}_k \} \mathbf{W}_i \right) }_{\text{BS distortion noise}} + \underbrace{ (1 + \rho_k) \sigma^2 }_{\text{Thermal noise}}.
\end{align}
From \eqref{eq141}, we observe that HIs introduce additional terms proportional to the transmit power and the channel gains, thus effectively increasing the noise floor and reducing the SINR and degrading the system's performance.
\subsection{Mitigation of HIs via FAS}
FAS offer a promising approach to mitigate the adverse effects of HIs by dynamically adjusting the antenna positions. By optimizing the positions of the transmit fluid antennas $\bar{\mathbf{t}}$ at the BS and the receive fluid antennas ${ \mathbf{r}_k }$ at the CUs, we can influence the channel characteristics and improve the SINR through the following mechanisms:
\subsubsection{Enhancement of Desired Signal Power} 
The desired signal power for the $k$-th CU is given by $\mathrm{Tr}\left( \mathbf{H}_k \mathbf{W}_k \right)$, which depends on the channel $\mathbf{h}_k$ as per \eqref{eq4}. Since $\mathbf{h}_k$ is a function of the antenna positions, optimizing $\bar{\mathbf{t}}$ and $\mathbf{r}_k$ can enhance the channel gain $| \mathbf{h}_k |^2$, thereby increasing the numerator of the SINR.

By strategically positioning the fluid antennas to align with the strongest propagation paths or to exploit favorable spatial channel conditions, the system can maximize the desired signal power received by each CU.
\subsubsection{Reduction of Inter-User Interference} 
 Inter-user interference, represented by the term $\sum_{i \neq k} \mathrm{Tr}\left( \mathbf{H}_k \mathbf{W}_i \right)$ in \eqref{eq14}. By adjusting the antenna positions, we can decorrelate the channels of different users, making them more orthogonal. This reduces the inter-user interference and enhances the SINR for each CU.
 
 Mathematically, we aim to minimize the cross-correlation between $\mathbf{h}_k$ and $\mathbf{h}_i$ for $i \neq k$, which can be quantified by the inner product $\mathbf{h}_k \mathbf{h}_i^H$. Optimizing antenna positions to minimize these inner products leads to reduced interference.
 
 \subsubsection{Mitigation of Distortion Noise Impact}
The distortion noise terms in $\delta_k$ are influenced by the channel gains and beamforming matrices, as seen in the terms involving $\mathrm{Tr}\left( \mathbf{H}_k \mathbf{W}_i \right)$ and $\mathrm{Tr}( \mathrm{diag}\{ \mathbf{H}_k \} \mathbf{W}_i )$. By optimizing the antenna positions to control these terms, we can mitigate the impact of HIs.

 \subsection{Problem Formulation}
 
 To quantitatively investigate the ability of FAS to mitigate HIs and enhance system performance, we formulate an optimization problem that aims to maximize the minimum SINR among all CUs. The optimization variables include the beamforming matrices $\{ \mathbf{W}_k \}_{k=1}^K$, the positions of the transmit antennas $\bar{\mathbf{t}}$, and the positions of the receive antennas $\{ \mathbf{r}_k \}_{k=1}^K$. The problem is constrained by the BS's transmit power limit, antenna position constraints, and physical limitations. Due to the monotonicity of the logarithmic function, the optimization problem can be formulated as
 \begin{subequations}\label{eq12}
 	\begin{align}
 		\max\limits_{\mathbf{\bar{t}},\{\mathbf{r}_k\}_{k=1}^K,\{\mathbf{W}_k\}_{k=1}^K}  \min\limits_{k \in \mathcal{K}}&	\qquad \gamma_k \label{eq12a}\\
 		\mathrm{s.t.} \quad \  &\mathbf{\overline{t}} \in \mathcal{S}_t, \label{eq12b}\\
 		&\mathbf{r}_k \in \mathcal{S}^k_r, k \in \mathcal{K}, \label{eq12c}\\
 		&||\mathbf{t}_n-\mathbf{t}_v||_2\geq D,~n,v\in\mathcal{N},~n\neq v, \label{eq12d}\\
 		&\sum\limits_{k=1}^{K}\mathrm{Tr}(\mathbf{W}_k) \leq P_{\max}, \label{eq12e}\\
 		&\mathrm{rank}(\mathbf{W}_k)=1, k \in \mathcal{K}, \label{eq12f}
 	\end{align}
 \end{subequations}
 where \eqref{eq12d} is the minimum distance requirement between the fluid antennas in the transmit region to avoid coupling, and $D$ is the predefined minimum distance between the transmit antennas; \eqref{eq12e} denotes the maximum transmit power constraint of the BS; \eqref{eq12f} is the rank-one constraint of $\{\mathbf{W}_k\}_{k=1}^K$. However, due to the highly non-convex objective function \eqref{eq12a}, constraint \eqref{eq12d}, and constraint \eqref{eq12f}, solving Problem \eqref{eq12} becomes difficult. To convexify the objective function \eqref{eq12a}, we introduce $K+1$ auxiliary variables $\tau$ and $\{\mu_k\}_{k=1}^K$, which results in the following optimization problem
 \begin{subequations}\label{eq13}
 	\begin{align}
 		\max\limits_{\mathbf{\bar{t}},\{\mathbf{r}_k\}_{k=1}^K,\{\mathbf{W}_k\}_{k=1}^K}& \tau \label{eq13a}\\
 		\mathrm{s.t.} \quad \ &\mathrm{Tr}\left(\mathbf{H}_k\mathbf{W}_k \right)\geq\tau \mu_k, k \in \mathcal{K},\label{eq13b}\\
 		&\mu_k \geq \delta_k, k \in \mathcal{K},\label{eq13c} \\
 		&\eqref{eq12b}-\eqref{eq12f}.
 	\end{align}
 \end{subequations}
Despite relaxing the above steps,  Problem \eqref{eq13} remains a non-convex optimization challenge due to the presence of non-convex constraints in \eqref{eq12d}, \eqref{eq12f}, \eqref{eq13b}, and \eqref{eq13c}. To address this complexity and thoroughly investigate the impact of FAS on the performance of multi-user system, we employ a BCD algorithm to slove Problem \eqref{eq13}. The detailed methodology of this approach is presented in the following section.
 
\section{BCD Algorithm}\label{sec:BCD}
In this section, we employ the BCD algorithm to decompose Problem \eqref{eq13} into three sub-problems, each of which is then reformulated into a convex form. By alternately optimizing these sub-problems, we iteratively converge to a locally optimal solution for Problem \eqref{eq13}.

\subsection{Optimization of BS's Transmit Beamforming}
Given $\mathbf{\bar{t}}$ and $\{\mathbf{r}_k\}_{k=1}^K$, the corresponding optimization problem of $\{\mathbf{W}_k\}_{k=1}^K$ can be expressed as
\begin{subequations}\label{eq14}
\begin{align}
\max\limits_{\{\mathbf{W}_k\}_{k=1}^K}& \tau \label{eq14a}\\
\mathrm{s.t.} \quad \ &\mathrm{Tr}\left(\mathbf{H}_k\mathbf{W}_k \right)\geq\tau \mu_k, k \in \mathcal{K},\label{eq14b}\\
&\mu_k \geq \delta_k, k \in \mathcal{K},\label{eq14c} \\
&\eqref{eq12e}, \eqref{eq12f}.
\end{align}
\end{subequations}
Problem \eqref{eq14} is a non-convex optimization problem due to non-convex constraints \eqref{eq12f} and \eqref{eq14c}.

For constraint \eqref{eq14b}, the product term $\tau$ and $\mu_k$ on the right-hand side of the inequality renders the constraint non-convex. To address this challenge, we apply an SCA approach to construct an upper bound for $\tau\mu_k$. Specifically, we utilize the identity
\begin{align}\label{eq15}
\tau\mu_k=\frac{1}{4}\left(\tau+\mu_k\right)^2-\frac{1}{4}\left(\tau-\mu_k\right)^2.
\end{align}
From \eqref{eq15}, we observe that both $\left(\tau+\mu_k\right)^2$ and $\left(\tau-\mu_k\right)^2$ are convex functions with respect to $\tau$ and $\mu_k$. Then, we obtain the upper bound of $\tau\mu_k$ by using the first-order Taylor expansion of $\left(\tau-\mu_k\right)^2$ at point $(\bar{\tau}, \bar{\mu_k})$, which is given by
\begin{align}\label{eq16}
f(\tau,\mu_k;\bar{\tau}, \bar{\mu_k})= & \frac{1}{4}\left(\tau+\mu_k\right)^2-\frac{1}{4}\left(\bar{\tau}-\bar{\mu}_k\right)^2 \nonumber \\
&-\frac{1}{2}\left(\bar{\tau}-\bar{\mu}_k\right)\left(\tau-\bar{\tau}-\mu_k+\bar{\mu}_k\right)
\geq \tau\mu_k.
\end{align}
For constraint \eqref{eq12f}, we employ the SRCR algorithm to reformulate the rank-one constraint equivalently as
\begin{align}\label{eq17}
\mathbf{u}_\text{max}^H\left(\mathbf{W}_k^{(p)}\right)\mathbf{W}_k\mathbf{u}_\text{max}\left( \mathbf{W}_k^{(p)}\right)\geq \vartheta^{(p)}\mathrm{Tr}(\mathbf{W}_k), k \in \mathcal{K},
\end{align}
where $\mathbf{W}_k^{(p)}$ represents $\mathbf{W}_k$ at the $p$-th internal iteration, $\mathbf{u}_\text{max}\left( \mathbf{W}_k^{(p)}\right)$ is the eigenvector associated with the largest eigenvalue $\lambda_\text{max}\left( \mathbf{W}_k^{(p)} \right)$ of $\mathbf{W}_k^{(p)}$ and $\vartheta^{(p)}$ represents the relaxation parameter and updated by the following relation
\begin{align}\label{eq18}
\vartheta^{(p)} = \min\left(1,\frac{\lambda_\text{max}\left( \mathbf{W}_k^{(p)}\right)}{\mathrm{Tr}(\mathbf{W}_k^{(p)})} + \alpha^{(p)} \right),
\end{align}
where $\alpha^{(p)}$ is a given scalar. This reformulation allows us to represent Problem \eqref{eq14} as
\begin{subequations}\label{eq19}
\begin{align}
\max\limits_{\{\mathbf{W}_k\}_{k=1}^K} &\tau \\
\mathrm{s.t.} \ \ &\mathrm{Tr}\left(\mathbf{H}_k\mathbf{W}_k \right) \geq f\left(\tau,\mu_k;\tau^{(p)}, \mu_k^{(p)}\right), k \in \mathcal{K},\\
&\eqref{eq12e},\eqref{eq14c}, \eqref{eq17},
\end{align}
\end{subequations}
where $\tau^{(p)}$ and $\{\mu_k^{(p)}\}_{k=1}^K$ represent the values of $\tau$ and $\{\mu_k\}_{k=1}^K$ at the $p$-th iteration. Since Problem \eqref{eq19} is now convex, it can be efficiently solved using a convex programming toolbox such as CVX \cite{MGrant}. We summarize the complete algorithmic process of Problem \eqref{eq14} in Algorithm 1.

\begin{algorithm}[h]
\caption{SCA-based SRCR Algorithm for Solving Problem \eqref{eq14}}
\begin{algorithmic}[1]
\STATE \textbf{Initialize:} $\tau^{(0)}$, $\{\mathbf{W}_k^{(0)}\}_{k=1}^K$, $\{\mu^{(0)}_k\}_{k=1}^K$, $\vartheta^{(0)}$, $\alpha^{(0)}$.
\STATE \textbf{while} Increase of the  auxiliary variables $\tau$ is above $\epsilon_w$ \textbf{do}\\
\STATE \quad \textbf{if}  If Problem \eqref{eq19} has a solution, then by solving \\ 
\quad\quad Problem \eqref{eq19}, we update $\{\mathbf{W}_k^{(p+1)}\}_{k=1}^K$ and letting \\
\quad\quad $\alpha^{(p+1)}=\alpha^0$ \\
\STATE \quad \textbf{else} \\
\STATE \quad \quad Update $\{\mathbf{W}_k^{(p+1)}\}_{k=1}^K=\{\mathbf{W}_k^{(p)}\}_{k=1}^K$;  \\
\STATE \quad \quad Update $\alpha^{(p+1)}=\alpha^{(p)}/2$;  \\
\STATE \quad \textbf{end if} \\
\STATE \quad Update $\vartheta^{(p+1)}$ by \eqref{eq18};
\STATE \quad $p:=p+1$;         \\
\STATE \textbf{end while}
\STATE Obtain $\{\mathbf{W}_k\}_{k=1}^K$, $\tau$, $\{\mu_k\}_{k=1}^K$.
\end{algorithmic}
\end{algorithm}

\subsection{Optimization of Fluid Antennas' Positions at the BS}
Given $\{\mathbf{W}_k\}_{k=1}^K$ and $\{\mathbf{r}_k\}_{k=1}^K$, Problem \eqref{eq13} can be formulated as
\begin{subequations}\label{eq20}
\begin{align}
\max\limits_{\mathbf{\overline{t}}} \ &\tau \label{eq20a}\\
\mathrm{s.t.} \ &\mathrm{Tr}\left(\mathbf{H}_k\mathbf{W}_k \right)\geq\tau \mu_k, k \in \mathcal{K},\label{eq20b}\\
&\mu_k \geq \delta_k, k \in \mathcal{K},\label{eq20c} \\
&\eqref{eq12b}, \eqref{eq12d}.
\end{align}
\end{subequations}
Problem \eqref{eq20} is a non-convex optimization problem due to non-convex constraints \eqref{eq20b}, \eqref{eq20c}, and \eqref{eq12d}.

For the right-hand side of constraint \eqref{eq20b}, we can obtain its upper bound, which is given by \eqref{eq16}. For the left-hand side of the inequality in constraint \eqref{eq20b}, we have
\begin{align}\label{eq21}
\mathrm{Tr}\left(\mathbf{H}_k\mathbf{W}_k \right)&=\mathrm{Tr}\left(\mathbf{f}^H_k(\mathbf{r}_k)\bm{\Sigma}_k\mathbf{G}_k\mathbf{(\bar{t})} \mathbf{w}_k \mathbf{w}_k^H
\mathbf{G}_k^H\mathbf{(\bar{t})} \bm{\Sigma}_k^H \mathbf{f}_k(\mathbf{r}_k) \right) \nonumber \\
&=\Upsilon_k(\mathbf{t}_n)+2\Re\{\mathbf{g}_k^H(\mathbf{t}_n) \bm{\beta}_k\}+\Lambda_k,
\end{align}
where
\begin{align}
\label{eq22}\Upsilon_k(\mathbf{t}_n)=&\mathrm{Tr}\left( \mathbf{w}_k(n)\mathbf{w}_k^H(n)\mathbf{g}_k(\mathbf{t}_n)
\mathbf{g}_k^H(\mathbf{t}_n)\bm{\Phi}_k  \right),\\
\label{eq23}\bm{\Phi}_k=&\bm{\Sigma}_k^H\mathbf{f}_k(\mathbf{r}_k)\mathbf{f}_k^H(\mathbf{r}_k)\bm{\Sigma}_k, \\
\label{eq24}\bm{\beta}_k=&\bm{\Phi}_k\left(\sum_{j\neq n}^{N} \mathbf{g}_k(\mathbf{t}_j)\mathbf{w}_k(j)\right)\mathbf{w}_k^H(n),\\
\label{eq25}\Lambda_k=&\mathrm{Tr}\left(\sum_{j\neq n}^{N} \mathbf{g}_k(\mathbf{t}_j)\mathbf{w}_k(j) \sum_{l\neq n}^{N} \mathbf{w}_k^H(l)\mathbf{g}_k^H(\mathbf{t}_l)\bm{\Phi}_k \right).
\end{align}
Hence, we turn to handle $\Upsilon_k(\mathbf{t}_n)$. First,  we provide a lower bound using the first-order Taylor expansion of  $\Upsilon_k(\mathbf{t}_n)$ at the point  $\mathbf{g}_k(\mathbf{t}_n^{(q)})$, which is given by
\begin{align}\label{eq26}
\Upsilon_k(\mathbf{t}_n) \geq &\mathbf{g}_k^H(\mathbf{t}_n^{(p)}) \bm{\tilde{\Phi}}_k \mathbf{g}_k(\mathbf{t}_n^{(p)}) \nonumber \\
&+2\Re\{ \mathbf{g}_k^H(\mathbf{t}_n^{(p)}) \bm{\tilde{\Phi}}_k (\mathbf{g}_k(\mathbf{t}_n) - \mathbf{g}_k(\mathbf{t}_n^{(p)}) )   \} \nonumber \\
=&2\Re\{ \mathbf{g}_k^H(\mathbf{t}_n^{(p)}) \bm{\tilde{\Phi}}_k \mathbf{g}_k(\mathbf{t}_n) \} - \omega_k,
\end{align}
where $\mathbf{t}_n^{(q)}$ is the $q$-th internal iteration of $\mathbf{t}_n$, $\bm{\tilde{\Phi}}_k=\bm{\Phi}_k\mathbf{w}_k(n)\mathbf{w}_k^H(n)$, and $\omega_k= \mathbf{g}_k^H(\mathbf{t}_n^{(q)}) \bm{\tilde{\Phi}}_k \mathbf{g}_k(\mathbf{t}_n^{(q)})$. We combine the second term in \eqref{eq21} with the first term in \eqref{eq26}, and the resulting new expression is
\begin{align}\label{eq27}
\tilde{\Upsilon}_k(\mathbf{t}_n)=2\Re\{\mathbf{g}_k^H(\mathbf{t}_n) \bm{\xi}_k\},
\end{align}
where $\bm{\xi}_k=\bm{\tilde{\Phi}}_k^H \mathbf{g}_k(\mathbf{t}_n^{(p)})+\bm{\beta}_k$. Although $\tilde{\Upsilon}_k(\mathbf{t}_n)$ is a linear function with respect to $\mathbf{g}_k(\mathbf{t}_n)$, it remains non-convex and non-concave with respect to $\mathbf{t}_n$. To address this issue, we use the second-order Taylor expansion to construct a surrogate function, which is the lower bound of $\mathrm{Tr}\left(\mathbf{H}_k\mathbf{W}_k \right)$. We denote the gradient vector and the Hessian matrix of $\tilde{\Upsilon}_k(\mathbf{t}_n)$ as $\nabla {\tilde{\Upsilon}_k}(\mathbf{t}_n)$ and $\nabla^2 {\tilde{\Upsilon}_k}(\mathbf{t}_n)$, respectively, whose detailed derivations are provided in Appendix \ref{appendixA}. Furthermore, we also introduce a scalar $\kappa_k^n$ such that $\kappa_k^n \mathbf{I}_2\succeq \nabla^2 {\tilde{\Upsilon}_k}(\mathbf{t}_n)$, whose detailed derivations are also provided in Appendix \ref{appendixA}. Thus, we can obtain the global lower bound of $\tilde{\Upsilon}_k(\mathbf{t}_n)$ as follows
\begin{align}\label{eq28}
{\hat{\Upsilon}_k}(\mathbf{t}_n)\triangleq &{\tilde{\Upsilon}_k}(\mathbf{t}_n^{(p)})+\nabla {\tilde{\Upsilon}_k}(\mathbf{t}_n^{(p)})^T\left(\mathbf{t}_n-\mathbf{t}_n^{(p)}   \right) \nonumber\\
&-\frac{\kappa_k^n}{2}\left(\mathbf{t}_n-\mathbf{t}_n^{(p)}   \right)^T\left(\mathbf{t}_n-\mathbf{t}_n^{(p)}\right)
\end{align}
From \eqref{eq28}, the non-convex constraint \eqref{eq20b} can be relaxed to
\begin{align}\label{eq29}
\hat{\Upsilon}_k(\mathbf{t}_n) + \Lambda_k - \omega_k \geq f\left(\tau,\mu_k;\tau^{(p)}, \mu_k^{(p)}\right).
\end{align}

From \eqref{eq20c}, we can find that $\delta_k$ is non-convex function with respect to $\mathbf{t}_n$. We first reformulate $\delta_k$ into
\begin{align}\label{eq30}
\delta_k=&(1+\rho_k)\sigma^2+ \mathrm{Tr}\left( \mathbf{G}_k\mathbf{(\bar{t})} \sum\limits_{i\neq k}^{K}\mathbf{w}_i \mathbf{w}_i^H  \mathbf{G}_k^H\mathbf{(\bar{t})} \bm{\Phi}_k \right)\nonumber\\
&+\rho_k\mathrm{Tr}\left( \mathbf{G}_k\mathbf{(\bar{t})} \sum\limits_{i=1}^{K}\mathbf{w}_i \mathbf{w}_i^H  \mathbf{G}_k^H\mathbf{(\bar{t})} \bm{\Phi}_k \right) \nonumber \\
&+(1+\rho_k) \eta \mathrm{Tr}\left(\mathbf{G}_k\mathbf{(\bar{t})} \text{diag}\left\{ \sum\limits_{i=1}^{K}\mathbf{w}_i \mathbf{w}_i^H\right\}\mathbf{G}_k^H\mathbf{(\bar{t})} \bm{\Phi}_k\right)\nonumber \\
=&\bm{\Xi}_k(\mathbf{t}_n)+2\Re\{\mathbf{g}_k^H(\mathbf{t}_n) \bm{\chi}_k\}+\Pi_k,
\end{align}
where
\begin{align}
\bm{\Xi}_k(\mathbf{t}_n)=&\mathrm{Tr}\left(\sum\limits_{i\neq k}^{K}\mathbf{w}_i(n)\mathbf{w}_i^H(n)\mathbf{g}_k(\mathbf{t}_n) \mathbf{g}_k^H(\mathbf{t}_n)\bm{\Phi}_k\right)\nonumber\\
&+(\rho_k+\eta+\rho_k\eta)\nonumber\\
&\cdot\mathrm{Tr}\left(\sum\limits_{i=1}^{K}\mathbf{w}_i(n)\mathbf{w}_i^H(n)\mathbf{g}_k(\mathbf{t}_n) \mathbf{g}_k^H(\mathbf{t}_n)\bm{\Phi}_k\right), \\
\bm{\chi}_k =&\sum\limits_{i\neq k}^{K} \bm{\Phi}_k\left(\sum_{j\neq n}^{N} \mathbf{g}_k(\mathbf{t}_j)\mathbf{w}_i(j)\right)\mathbf{w}_i^H(n) \nonumber\\
&+\rho_k\sum\limits_{i=1}^{K} \bm{\Phi}_k\left(\sum_{j\neq n}^{N} \mathbf{g}_k(\mathbf{t}_j)\mathbf{w}_i(j)\right)\mathbf{w}_i^H(n),
\end{align}
\begin{align}
\Pi_k =&\sum\limits_{i\neq k}^{K}\left( \mathrm{Tr}\left(\sum_{j\neq n}^{N} \mathbf{g}_k(\mathbf{t}_j)\mathbf{w}_i(j) \sum_{l\neq n}^{N} \mathbf{w}_i^H(l)\mathbf{g}_k^H(\mathbf{t}_l)\bm{\Phi}_k \right) \right)\nonumber\\
&+\rho_k\sum\limits_{i=1}^{K}\left( \mathrm{Tr}\left(\sum_{j\neq n}^{N} \mathbf{g}_k(\mathbf{t}_j)\mathbf{w}_i(j) \sum_{l\neq n}^{N} \mathbf{w}_i^H(l)\mathbf{g}_k^H(\mathbf{t}_l)\bm{\Phi}_k \right) \right)\nonumber\\
&+(1+\rho_k) \eta\sum\limits_{i=1}^{K} \mathrm{Tr}\left( \sum\limits_{j \neq n}^{N} \mathbf{g}_k(\mathbf{t}_j) \mathbf{w}_i(j)\mathbf{w}_i^H(j) \mathbf{g}_k^H(\mathbf{t}_j)\bm{\Phi}_k\right)\nonumber\\
&+(1+\rho_k)\sigma^2.
\end{align}

Then, we employ the MM algorithm to further address $\bm{\Xi}_k(\mathbf{t}_n)$. According to \cite{JTang2024}, for a given $\mathbf{g}_k^H(\mathbf{t}_n^{(m)})$, the following inequality holds for any feasible $\mathbf{g}_k^H(\mathbf{t}_n)$, i.e.,\begin{align}\label{eq37}
\bm{\Xi}_k(\mathbf{t}_n) \leq &\mathbf{g}_k^H(\mathbf{t}_n) \bm{\Theta}_k \mathbf{g}_k(\mathbf{t}_n) + \mathbf{g}_k^H(\mathbf{t}_n^{(p)}) \left( \bm{\Theta}_k - \bm{\hat{\Phi}}_k \right) \mathbf{g}_k(\mathbf{t}_n^{(p)}) \nonumber \\
&-2\Re\left(  \mathbf{g}_k^H(\mathbf{t}_n) \left( \bm{\Theta}_k - \bm{\hat{\Phi}}_k \right) \mathbf{g}_k(\mathbf{t}_n^{(p)})         \right),
\end{align}
where
\begin{align}
\label{eq38}\bm{\hat{\Phi}}_k=&\bm{\Phi}_k \sum\limits_{i\neq k}^{K} \mathbf{w}_i(n)\mathbf{w}_i^H(n)\nonumber\\
&+(\rho_k+\eta+\rho_k\eta)\bm{\Phi}_k \sum\limits_{i=1}^{K} \mathbf{w}_i(n)\mathbf{w}_i^H(n),\\
\label{eq39}\bm{\Theta}_k=&\lambda_\text{max}\mathbf{I}_{L_k^t},
\end{align}
and $\lambda_\text{max}$ is the maximum eigenvalue of $\bm{\hat{\Phi}}_k$. From \eqref{eq37}, we know that $\mathbf{g}_k^H(\mathbf{t}_n) \bm{\Theta}_k \mathbf{g}_k(\mathbf{t}_n)=\lambda_\text{max} L_k^t$, and the second term is constant. We define
\begin{align}\label{eq40}
C_k\triangleq\lambda_\text{max} L_k^t+\mathbf{g}_k^H(\mathbf{t}_n^{(p)}) \left( \bm{\Theta}_k - \bm{\hat{\Phi}}_k \right) \mathbf{g}_k(\mathbf{t}_n^{(p)}).
\end{align}
Thus, an upper bound on $\delta_k$ can be written as
\begin{align}\label{eq41}
\underbrace{2\Re\{\mathbf{g}_k^H(\mathbf{t}_n) \bm{\zeta}_k\}}_{{\bm{\hat{\Xi}}_k}(\mathbf{t}_n)}+C_k+\Pi_k,
\end{align}
where $\bm{\zeta}_k=\bm{\chi}_k- \left( \bm{\Theta}_k - \bm{\hat{\Phi}}_k \right) \mathbf{g}_k(\mathbf{t}_n^{(p)})$. However, the concavity of ${\bm{\hat{\Xi}}_k}(\mathbf{t}_n)$ with respect to $\mathbf{t}_n$ is still not determined. We rewrite $\bm{\hat{\Xi}}_k$ as
\begin{align}
\bm{\hat{\Xi}}_k=2\left( \sum_{l=1 }^{L_k^t}   \lvert   \bm{\zeta}_k^{l}   \rvert  \cos\left(\nu_k^{l}(\mathbf{t}_n)\right)\right),
\end{align}
where
\begin{align}
\nu_k^{l}(\mathbf{t}_n)=\frac{2\pi}{\lambda}\varrho_{k,l}^t(\mathbf{t}_n)-\angle\bm{\zeta}_k^{l}.
\end{align}

As shown in Appendix \ref{appendixA}, we can obtain the gradient vector and the Hessian matrix of $\bm{\hat{\Xi}}_k$ as $\nabla \bm{\hat{\Xi}}_k$ and $\nabla^2 \bm{\hat{\Xi}}_k$, and a scalar $\tilde{\kappa}_k^n$.
Then, we also utilize the second-order Taylor expansion to construct the upper bound of ${\bm{\hat{\Xi}}_k}(\mathbf{t}_n)$, which is
\begin{align}\label{eq42}
\bm{\tilde{\Xi}}_k(\mathbf{t}_n)= &\bm{\hat{\Xi}}_k (\mathbf{t}_n^{(p)})+\nabla \bm{\hat{\Xi}}_k(\mathbf{t}_n^{(p)})^T\left(\mathbf{t}_n-\mathbf{t}_n^{(p)}\right) \nonumber\\
&+\frac{\tilde{\kappa}_k^n}{2}\left(\mathbf{t}_n-\mathbf{t}_n^{(p)}\right)^T\left(\mathbf{t}_n-\mathbf{t}_n^{(p)}\right).
\end{align}

We can obtain an upper bound on $\delta_k$ as follow
\begin{align}\label{eq48}
f_k^u(\mathbf{t}_n) = \bm{\tilde{\Xi}}_k(\mathbf{t}_n)+C_k+\Pi_k.
\end{align}
Therefore, constraint \eqref{eq20c} can be relaxed as
\begin{align}\label{eq49}
f_k^u(\mathbf{t}_n) \leq \mu_k, k \in \mathcal{K}.
\end{align}

For the constraint \eqref{eq12d}, we can relax $||\mathbf{t}_n-\mathbf{t}_v||_2$ as a concave function of $\mathbf{t}_n$ to its lower bound by using the first-order Taylor expansion at point $\mathbf{t}_n^{(q)}$. Then, we have
\begin{align}\label{eq50}
||\mathbf{t}_n-\mathbf{t}_v||_2 &\geq ||\mathbf{t}_n^{(p)}-\mathbf{t}_v||_2+\left(\nabla ||\mathbf{t}_n^{(p)}-\mathbf{t}_v||_2\right)^T \left(\mathbf{t}_n-\mathbf{t}_n^{(p)} \right) \nonumber \\
&=\frac{1}{||\mathbf{t}_n^{(p)}-\mathbf{t}_v||_2}(\mathbf{t}_n^{(p)}-\mathbf{t}_v)^T(\mathbf{t}_n-\mathbf{t}_v).
\end{align}
Therefore, constraint \eqref{eq12d} can be written as
\begin{align}\label{eq51}
\frac{1}{||\mathbf{t}_n^{(p)}-\mathbf{t}_v||_2}(\mathbf{t}_n^{(p)}-\mathbf{t}_v)^T(\mathbf{t}_n-\mathbf{t}_v) \geq D.
\end{align}

Based on the above-mentioned derivations, Problem \eqref{eq20} can be reformulated as
\begin{subequations}\label{eq52}
\begin{align}
\max\limits_{\mathbf{\bar{t}}} \quad &\tau \label{eq52a}\\
\mathrm{s.t.} \quad  &\eqref{eq12b},\eqref{eq29},\eqref{eq49},\eqref{eq51}.
\end{align}
\end{subequations}
Problem \eqref{eq52} is convex, it can be efficiently solved using the convex programming toolbox CVX \cite{MGrant}.

\subsection{Optimization of Fluid Antennas' Positions at the CUs}
Given $\{\mathbf{W}_k\}_{k=1}^K$ and $\mathbf{\bar{t}}$, Problem \eqref{eq13} can be rewritten as
\begin{subequations}\label{eq53}
\begin{align}
\max\limits_{\{\mathbf{r}_k\}_{k=1}^K} \ &\tau \label{eq20a}\\
\mathrm{s.t.} \ &\mathrm{Tr}\left(\mathbf{H}_k\mathbf{W}_k \right)\geq\tau \mu_k, k \in \mathcal{K},\label{eq53b}\\
&\mu_k \geq \delta_k, k \in \mathcal{K},\label{eq53c} \\
&\eqref{eq12c}.
\end{align}
\end{subequations}
Problem \eqref{eq53} is a non-convex optimization problem due to non-convex constraints \eqref{eq53b} and \eqref{eq53c}.

For the right-hand side of the constraint \eqref{eq53b}, the upper bound as $\tau \mu_k$ is given in \eqref{eq16}. For the left-hand side, we have
\begin{align}\label{eq54}
\mathrm{Tr}\left(\mathbf{H}_k\mathbf{W}_k \right) = \mathbf{f}^H_k(\mathbf{r}_k)\mathbf{A}_k \mathbf{f}_k(\mathbf{r}_k),
\end{align}
where
\begin{align}\label{eq55}
\mathbf{A}_k = \bm{\Sigma}_k\mathbf{G}_k\mathbf{(\overline{t})} \mathbf{w}_k \mathbf{w}_k^H \mathbf{G}_k^H\mathbf{(\overline{t})} \bm{\Sigma}_k^H.
\end{align}
We can find that the structure of \eqref{eq54} is similar to that of $\Upsilon_k(\mathbf{t}_n)$. Therefore, we can obtain the lower bound of $\mathrm{Tr}\left(\mathbf{H}_k\mathbf{W}_k \right)$ using a similar approach, which can be expressed as
\begin{align}\label{eq56}
g_k^l(\mathbf{r}_k)={\hat{\Gamma}_k}(\mathbf{r}_k) - \varpi_k.
\end{align}
where
\begin{align}\label{eq57}
{\hat{\Gamma}_k}(\mathbf{t}_k)= &{{\Gamma}_k}(\mathbf{r}_k^{(p)})+\nabla {{\Gamma}_k}(\mathbf{r}_k^{(p)})^T\left(\mathbf{r}_k-\mathbf{r}_k^{(p)}\right) \nonumber\\
&-\frac{\hat{\kappa}_k^n}{2}\left(\mathbf{r}_k-\mathbf{r}_k^{(p)}\right)^T\left(\mathbf{r}_k-\mathbf{r}_k^{(p)}\right),\\
{\Gamma}_k(\mathbf{t}_n)=&2\Re\{\mathbf{f}_k^H(\mathbf{r}_k) \mathbf{A}_k^H \mathbf{f}_k(\mathbf{r}_k^{(p)})\},\\
\varpi_k= &\mathbf{f}_k^H(\mathbf{r}_k^{(p)}) \mathbf{A}_k \mathbf{f}_k(\mathbf{r}_k^{(p)}).
\end{align}
Furthermore, the derivations of $\nabla {\Gamma}_k(\mathbf{r}_k^{(p)})^T$ and $\hat{\kappa}_k^n$ are similar to the derivations provided in Appendix \ref{appendixA}, as such are omitted.

For the constraint \eqref{eq53c}, $\delta_k$ can be rewritten as
\begin{align}\label{eq58}
\delta_k =& \mathbf{f}^H_k(\mathbf{r}_k)\mathbf{B}_k\mathbf{f}_k(\mathbf{r}_k)+(1+\rho_k)\sigma^2,
\end{align}
where
\begin{align}
\mathbf{B}_k=&\bm{\Sigma}_k\mathbf{G}_k\mathbf{(\bar{t})} \sum\limits_{i\neq k}^{K}\mathbf{w}_i \mathbf{w}_i^H \mathbf{G}_k^H\mathbf{(\bar{t})} \bm{\Sigma}_k^H \nonumber\\
&+\rho_k\bm{\Sigma}_k\mathbf{G}_k\mathbf{(\bar{t})} \sum\limits_{i=1}^{K}\mathbf{w}_i \mathbf{w}_i^H \mathbf{G}_k^H\mathbf{(\bar{t})} \bm{\Sigma}_k^H \nonumber\\
&+(1+\rho_k) \eta \bm{\Sigma}_k\mathbf{G}_k\text{diag}\left\{\sum\limits_{i=1}^{K} \mathbf{w}_i \mathbf{w}_i^H\right\}\mathbf{G}_k^H\mathbf{(\bar{t})} \bm{\Sigma}_k^H.
\end{align}
Similar to \eqref{eq30}, we can   use the MM algorithm to obtain the upper bound of $\delta_k$, which is given by
\begin{align}\label{eq59}
g_k^u(\mathbf{r}_k)=&{\check{\Gamma}_k}(\mathbf{r}_k)+(1+\rho_k)\sigma^2\nonumber\\
&+\lambda^r_\text{max} L_k^t+\mathbf{f}_k^H(\mathbf{r}_k^{(p)}) \left( \mathbf{C}_k - \mathbf{B}_k \right)\mathbf{f}_k(\mathbf{r}_k^{(p)}).
\end{align}
where
\begin{align}
{\check{\Gamma}_k}(\mathbf{t}_k)= &{\tilde{\Gamma}_k}(\mathbf{r}_k^{(p)})+\nabla {\tilde{\Gamma}_k}(\mathbf{r}_k^{(p)})^T\left(\mathbf{r}_k-\mathbf{r}_k^{(p)}\right) \nonumber\\
&+\frac{\check{\kappa}_k^n}{2}\left(\mathbf{r}_k-\mathbf{r}_k^{(p)}\right)^T\left(\mathbf{r}_k-\mathbf{r}_k^{(p)}\right),\\
\tilde{\Gamma}_k(\mathbf{t}_n)=&2\Re\{\mathbf{f}_k^H(\mathbf{r}_k) (\mathbf{B}_k-\mathbf{C}_k) \mathbf{f}_k(\mathbf{r}_k^{(p)})\},\\
\mathbf{C}_k =& \lambda^r_\text{max}\mathbf{I}_{L_k^r},
\end{align}
and $\lambda^r_\text{max}$ is the maximum eigenvalue of $\mathbf{B}_k$. Furthermore, the derivations of $\nabla {\check{\Gamma}_k}(\mathbf{r}_k^{(p)})^T$ and $\check{\kappa}_k^n$ are similar to those in Appendix \ref{appendixA}, and as such are omitted.

Base on the above-mentioned derivations, Problem \eqref{eq53} can be reformulated as
\begin{subequations}\label{eq63}
\begin{align}
\max\limits_{\{\mathbf{r}_k\}_{k=1}^K} \quad &\tau \label{eq63a}\\
\mathrm{s.t.} \quad &g_k^l(\mathbf{r}_k)\geq f\left(\tau,\mu_k;\tau^{(p)}, \mu_k^{(p)}\right), k \in \mathcal{K},\label{eq63b}\\
&\mu_k \geq g_k^u(\mathbf{r}_k), k \in \mathcal{K},\label{eq63c} \\
&\eqref{eq12c}.
\end{align}
\end{subequations}
Problem \eqref{eq63} is convex, it can be efficiently solved using the convex programming toolbox CVX \cite{MGrant}.

Finally, we summarize the complete algorithmic process in Algorithm 2.

\begin{algorithm}[h]
\caption{Block Coordinate Descent Algorithm for Solving Problem \eqref{eq13}}
\begin{algorithmic}[1]
\STATE \textbf{Initialize:} $\tau^{(0)}$, $\{\mathbf{W}_k^{(0)}\}_{k=1}^K$, $\{\mu^{(0)}_k\}_{k=1}^K$, $\mathbf{\bar{t}}^{(0)}$, $\mathbf{r}_k^{(0)}$.
\STATE \textbf{while} Increase of the  auxiliary variables $\tau$ is above $\epsilon$ \textbf{do} \\ 
\STATE \quad \textbf{while} Increase of the  auxiliary variables $\tau$ is above $\epsilon_w$  \\ 
\quad \textbf{do}
\STATE \quad \quad Update $\{\mathbf{W}_k\}_{k=1}^K$ by solving Problem \eqref{eq19};
\STATE \quad \textbf{end while}
\STATE \quad \textbf{while} Increase of the  auxiliary variables $\tau$ is above $\epsilon_t$  \\ 
\quad \textbf{do}
\STATE \quad \quad \textbf{for} $n=1\rightarrow N$ \textbf{do} \\
\STATE \quad \quad \quad Update $\mathbf{\bar{t}}$ by solving Problem \eqref{eq52};
\STATE \quad \quad \textbf{end}
\STATE \quad \textbf{end while}
\STATE \quad \textbf{while} Increase of the  auxiliary variables $\tau$ is above $\epsilon_r$  \\ 
\quad \textbf{do}
\STATE \quad \quad Update $\{\mathbf{r}_k\}_{k=1}^K$ by solving Problem  \eqref{eq63};
\STATE \quad \textbf{end while}
\STATE \textbf{end while}
\STATE Obtain $\mathbf{W}_k$, $\mathbf{\bar{t}}$, $\mathbf{r}_k$.
\end{algorithmic}
\end{algorithm}

\emph{Complexity Analysis}: The computational complexity of solving an semidefinite programming (SDP) problem is
$\mathcal{O}\left( n_{\text{sdp}}^{0.5}(m_{\text{sdp}}n_{\text{sdp}}^{3}+m_{\text{sdp}}^{2}n_{\text{sdp}}^{2}+m_{\text{sdp}}^{3}) \ln(1/\delta)  \right)$, where $m_{\text{sdp}}$ is the number of semidefinite cone constraints, $n_{\text{sdp}}$ is the dimension of the semidefinite cone, and $\delta$ denotes the accuracy \cite{SBoyd}. Problem \eqref{eq19} is an SDP problem, $m_{\text{sdp}}=2K+1$ and $n_{\text{sdp}}=N$. Therefore, the computation complexity of Problem \eqref{eq19}  by using Algorithm 1 is
\begin{align} 
\mathcal{O}\left( L_1((2K+1)N^{3.5}+(2K+1)^2N^{2.5}\right.\nonumber\\
\left.+(2K+1)^3N^{0.5}) \ln(1/\delta)  \right),
\end{align}
where $L_1$ represents the number of iterations required for the convergence of Algorithm 1.

For Problem \eqref{eq52}, the computational complexities of $\bm{\zeta}_k$ in \eqref{eq41} is $\mathcal{O}\left((L^k_{k})^2\right)$; The computational complexities of $\nabla \bm{\hat{\Xi}}_k(\mathbf{t}_n^{(p)})$ and $\tilde{\kappa}_k^n$ in \eqref{eq42} and \eqref{eq42} are $\mathcal{O}\left(L^t_{k}\right)$ and $\mathcal{O}\left(1\right)$, respectively; The computational complexity of solving Problem \eqref{eq52} in each iteration is $\mathcal{O}\left((N+K)^{1.5}\ln(1/\delta)\right)$. Thus, the computational complexity of solving Problem \eqref{eq52} is $\mathcal{O}\left( NL_2 \sum_{k=1}^K \left((L^k_{k})^2 +(N+K)^{1.5}\ln(1/\delta)\right) \right)$, where $L_2$ is the number of iterations required for the convergence of Problem \eqref{eq52}. For the Problem \eqref{eq63}, we can obtain the complexity for optimizing the receive fluid antennas as $\mathcal{O}\left(  NL_3\sum_{k=1}^K((L^r_{k})^2 + (2K)^{1.5}\ln(1/\delta) ) \right)$, where $L_3$ is the number of iterations required for the convergence of Problem \eqref{eq63}. Therefore, the computational complexity of each iteration of Algorithm 2 is
\begin{align}
&\mathcal{O}\left( L_1((2K+1)N^{3.5}+(2K+1)^2N^{2.5}+(2K+1)^3N^{0.5})\right.\nonumber\\
&\left. \ln(1/\delta)  +NL_2 \sum_{k=1}^K \left((L^k_{k})^2 +(N+K)^{1.5}\ln(1/\delta)\right)\right.\nonumber\\
&\left. NL_3\sum_{k=1}^K((L^r_{k})^2 + (2K)^{1.5}\ln(1/\delta) ) \right).
\end{align}

\section{Numerical Result}\label{sec:Num}

In our simulations, we assume that the BS is located at (0, 0) m, and the number of users is 2, i.e., $K=2$, which are randomly distributed within the rectangular area from (20m, 0m) to (40m, -20m). The number of BS's fluid antennas is $N=4$, while each user is equipped with a single fluid antenna. The carrier frequency is set to be 2.4 GHz, resulting in a wavelength of $\lambda$ = 0.125 m, and the minimum inter-antenna distance is $D=\frac{\lambda}{2}$. We also assume that the moving region at the BS and users are both $S_t=S_r^k=\left[-{\frac{A}{2}},{\frac{A}{2}}  \right] \times \left[-{\frac{A}{2}},{\frac{A}{2}}  \right]$, where $A=4\lambda$. In the geometric channel, for $k$-th user, the number of transmit paths equals the number of receive paths, i.e., $L_k^t=L_k^r=L=5$. Meanwhile, both the AoDs and AoAs at the BS and users are independent and identically distributed variables, such that $\theta_{k,l}^t\sim\mathcal{U}[0,\pi], \varphi_{k,l}^t\sim\mathcal{U}[0,\pi], \theta_{k,m}^r\sim\mathcal{U}[0,\pi], \varphi_{k,m}^r\sim\mathcal{U}[0,\pi]$. Furthermore, we model the path response matrix for non-line-of-sight (NLoS) links between the BS and users as $\bm{\Sigma}_k[l,l] \sim \mathcal{CN}(0,g_0 \left( d_k/d_0\right)^{-\alpha}/L), l=1,2,\cdots, L$, where $d_k$ is the distance between the BS and the $k$-th user, $g_0=-30$ dB represents the average channel gain at the reference distance $d_0=1$m, and $\alpha=2.8$ is the path loss exponent \cite{LZhou24,TWu20241}. We also assume that the maximum transmit power of the BS is $P_\text{max}=30$ dBm, the noise power at all users is $\sigma_1^2=\sigma^2_2=\cdots=-80$ dBm \cite{JYao123}. Furthermore, the hardware degradation factor at the BS equals to those at the users, i.e., $\eta=\rho_1=\rho_2=\cdots=\rho_K$. If there are no specific requirements, both of these values are set as 0.2. The simulation results are averaged over 100 randomly generated channel realizations and the convergence accuracy is $10^{-3}$.

\begin{figure}[t]
\centering
\includegraphics[width=3.5in]{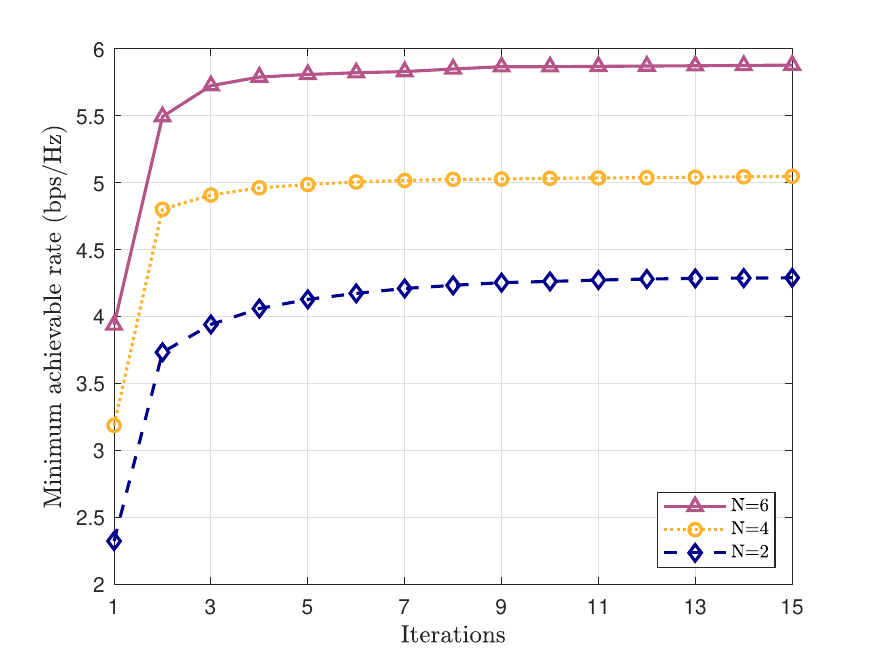}
\caption { Convergence of the proposed BCD algorithm.}
\label{covergence}
\end{figure}
In Fig.~\ref{covergence}, we present the convergence behavior of our employed BCD algorithm for a BS equipped with $N=2$, $4$, and $6$ fluid antennas, respectively. As shown, the algorithm converges after approximately 11 iterations in all three cases, demonstrating rapid convergence and confirming its effectiveness. Furthermore, we observe that the achievable rate increases as the number of fluid antennas increases.

\begin{figure}[t]
\centering
\includegraphics[width=3.5in]{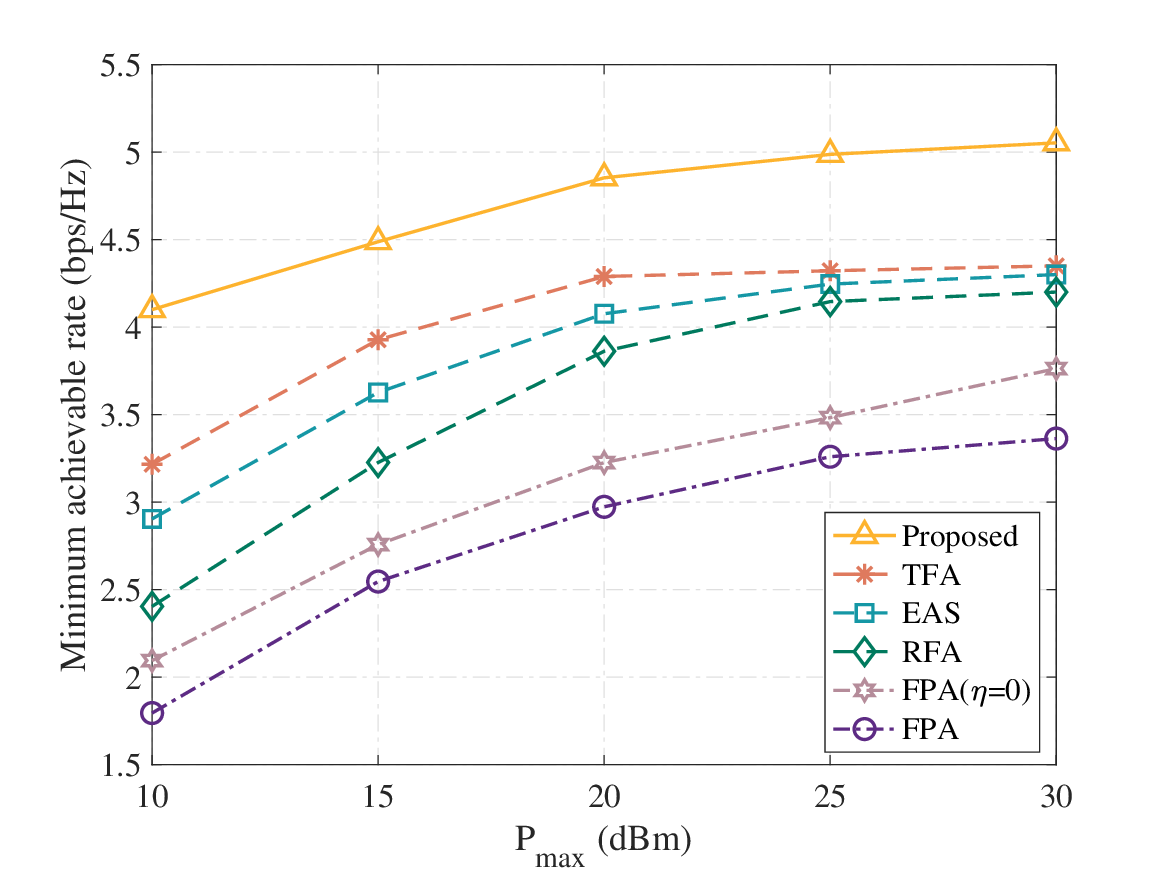}
\caption {The maximum transmit power of the BS $P_{\max}$ versus the minimum achievable rate.}
\label{RvsP}
\end{figure}

\begin{figure}[t]
\centering
\includegraphics[width=3.5in]{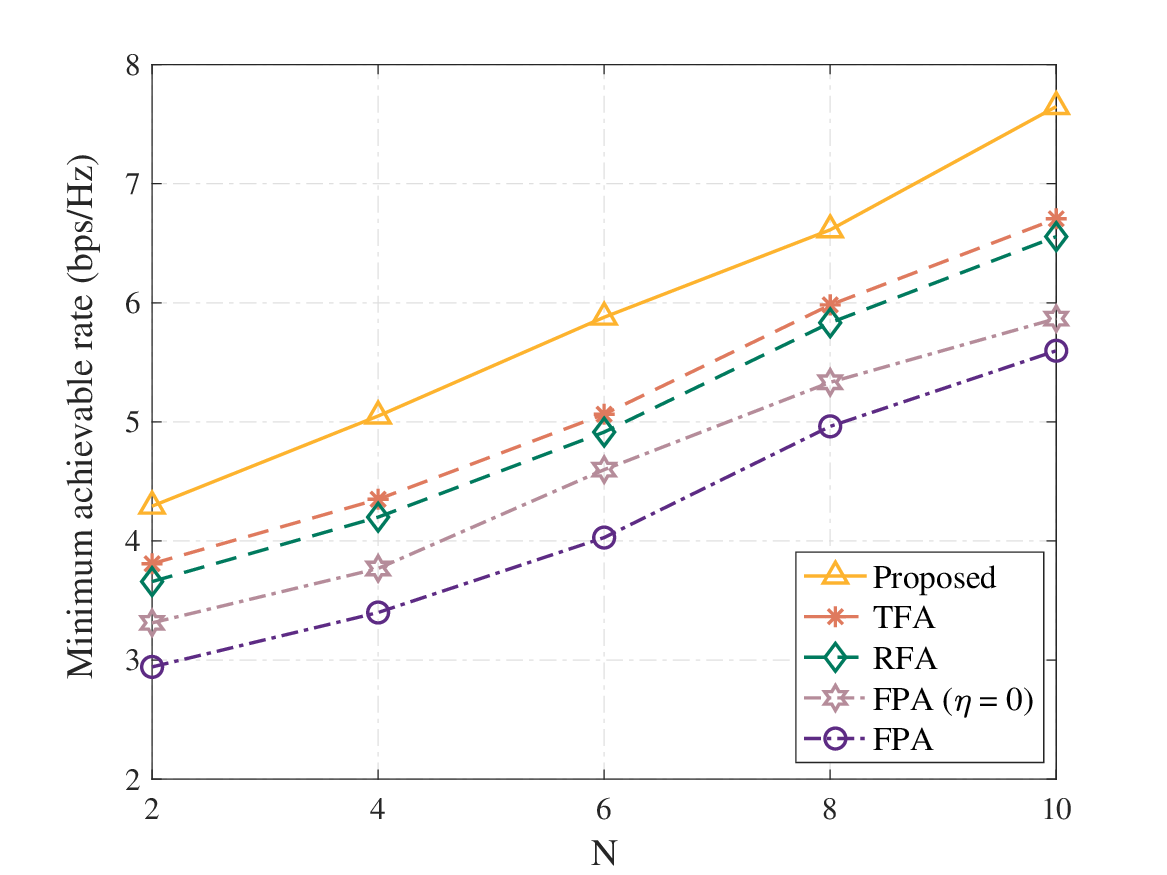}
\caption {The number of fluid antennas at the BS $N$ versus the minimum achievable rate.}
\label{RvsN}
\end{figure}

\begin{figure}[t]
\centering
\includegraphics[width=3.3in]{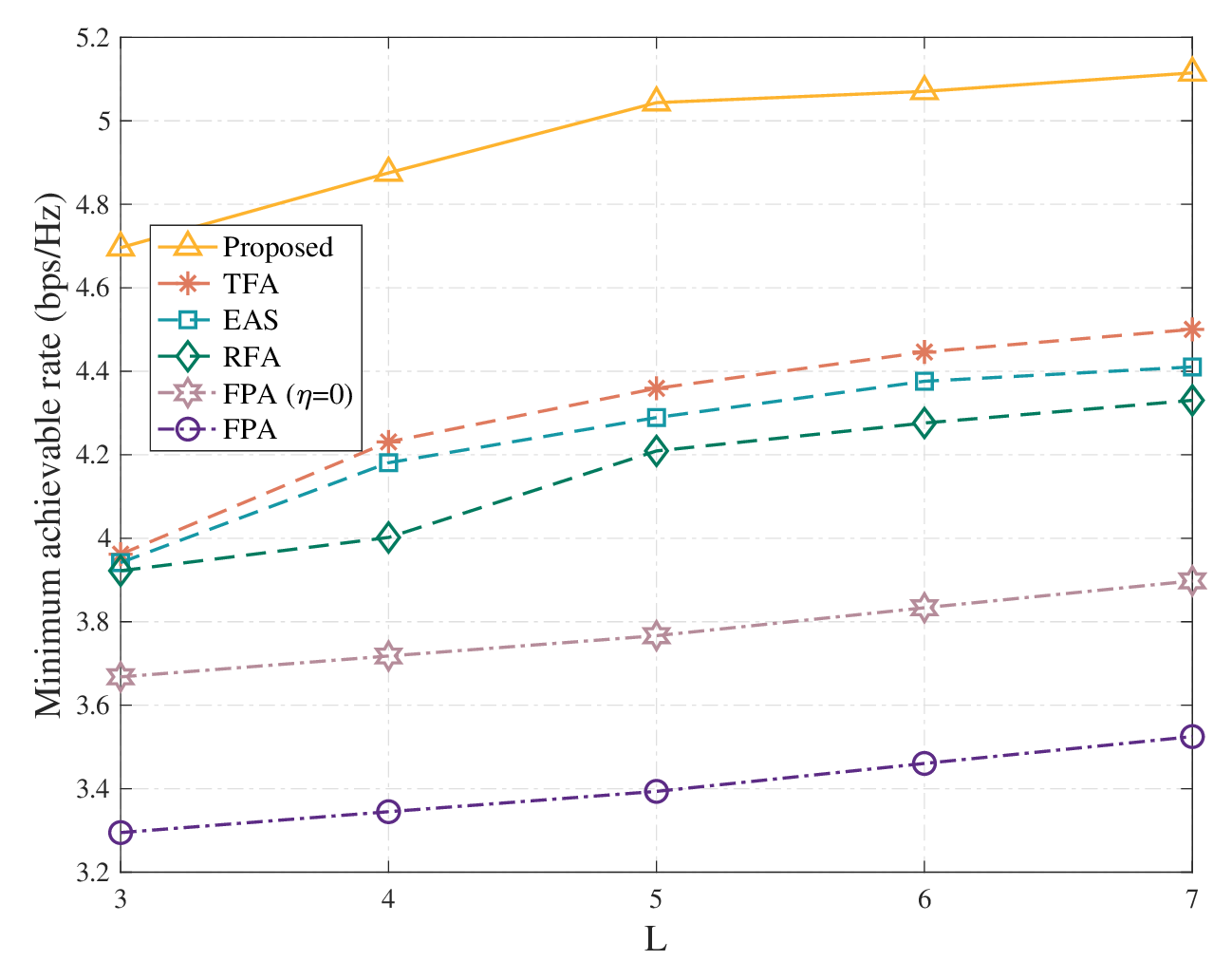}
\caption {The number of path $L$ versus the minimum achievable rate.}
\label{RvsL}
\end{figure}

\begin{figure}[t]
\centering
\includegraphics[width=3.5in]{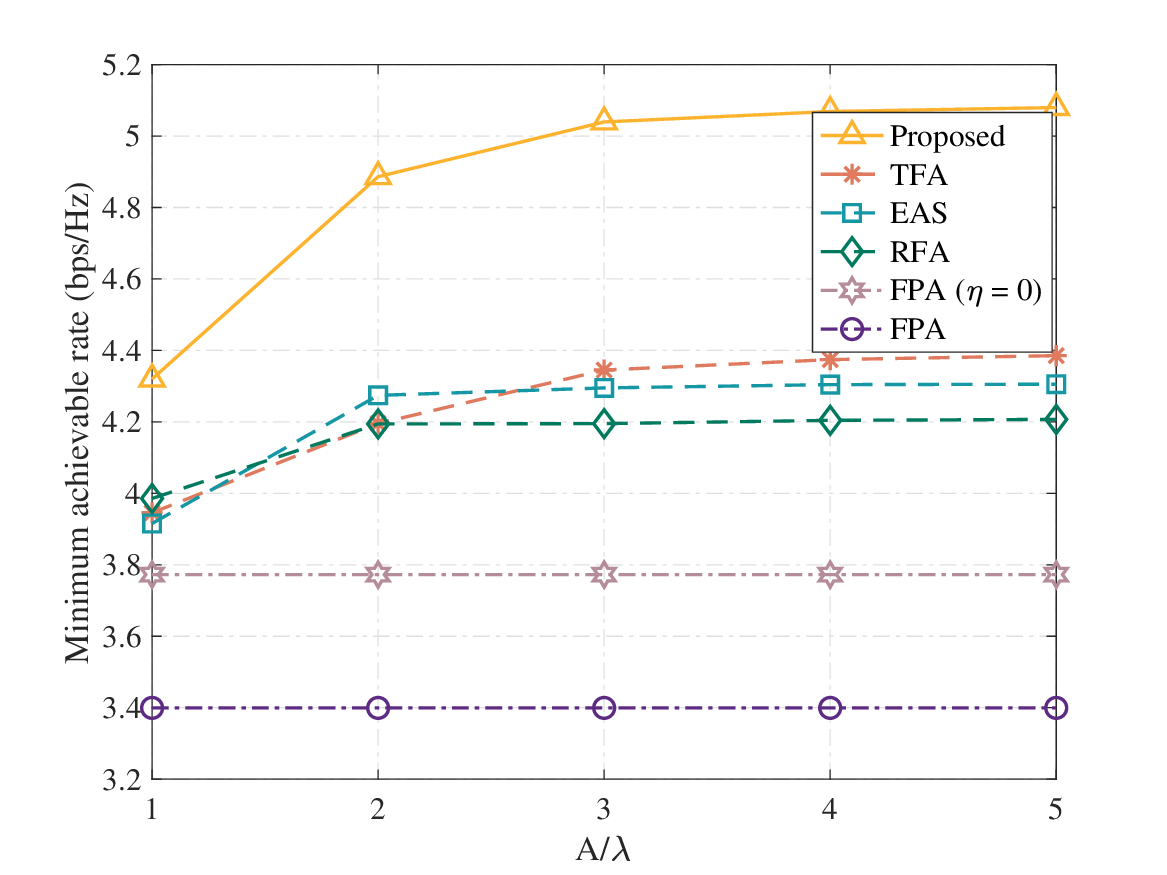}
\caption {The moving region $A/\lambda$ versus the minimum achievable rate.}
\label{RvsA}
\end{figure}

\begin{figure}[t]
\centering
\includegraphics[width=3.5in]{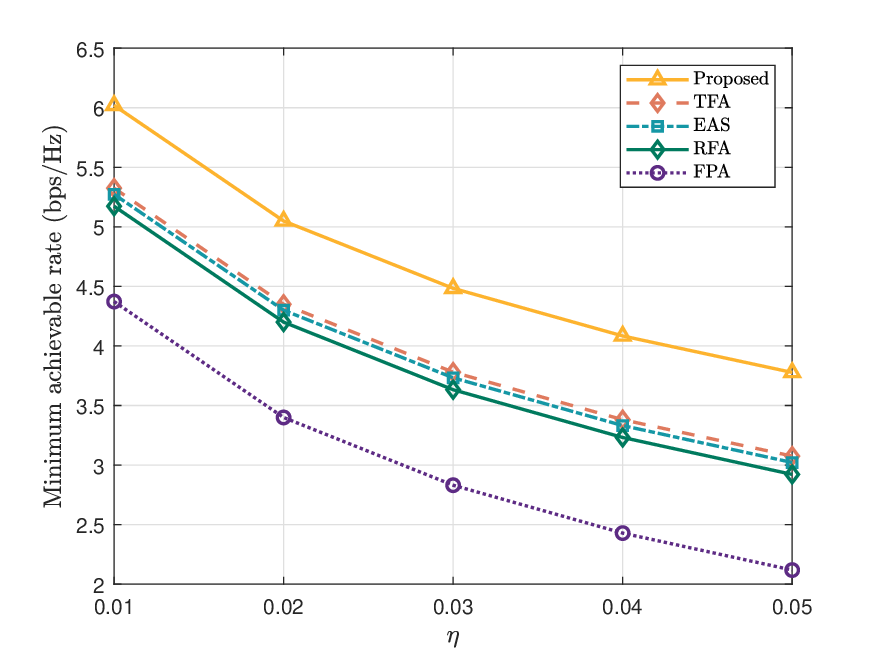}
\caption {The hardware degradation factor $\eta$ versus the minimum achievable rate.}
\label{Rvseta}
\end{figure}

In Fig.~\ref{RvsP}, Fig.~\ref{RvsN}, Fig.~\ref{RvsL}, Fig.~\ref{RvsA}, and Fig.~\ref{Rvseta}, we compare the proposed scheme with the following benchmarks:

$\mathbf{TFA}$: The BS is equipped with $N$ fluid antennas, while CUs are all equipped with a single FPA.

$\mathbf{EAS}$: The BS is equipped with an FPA-based UPA, consisting of $2N$ antennas, from which $N$ antennas are selected through an exhaustive search.

$\mathbf{RFA}$: The BS is equipped with $N$ FPAs, while CUs are all equipped with a single  fluid atnenna.

$\mathbf{FPA}$: The BS is equipped with $N$ FPAs, while UE is equipped with a single FPA.
  
In Fig.~\ref{RvsP}, we compare the minimum achievable rate $R$ of various schemes under different maximum transmit power at the BS $P_{\max}$. It is observed that as the maximum transmit power of the BS increases, the achievable rate of all schemes increases. However, when $P_{\max}$ exceeds $20$ dBm, the improvements become less significant. We infer that this behavior is due to two factors: \textit{first, HIs become more pronounced at higher power levels, which diminishes the benefits of increased transmit power; second, the achievable rate approaches its theoretical upper limit as power increases, resulting in diminishing returns despite further power increase.}

Furthermore, our results indicate that FAS can significantly enhance communication performance in the presence of HIs, with the minimum achievable rate increasing by approximately $128\%$. \textit{This demonstrates the effectiveness of FAS in mitigating the adverse effects of HIs.} Notably, the best performance is achieved when both the BS and the CUs are equipped with fluid antennas, highlighting the substantial benefits of deploying FAS on both ends of the communication link. This configuration not only maximizes the achievable rate but also enhances system robustness under hardware constraints. Additionally, equipping the BS with fluid antennas allows for significant improvements in the minimum achievable rate, particularly under low transmit power conditions.

Moreover, the results reveal that the best performance is achieved when both the BS and CUs are equipped with FAS, as this configuration maximizes the achievable rate and enhances system robustness against HIs. Notably, equipping only the BS with FAS also delivers significant improvements in the minimum achievable rate, particularly under low transmit power conditions. These findings reaffirm the potential of FAS in mitigating hardware constraints and optimizing multi-user system performance, directly confirming that \textit{\textbf{FAS makes a measurable difference in alleviating the impact of HIs.}}

\textbf{\emph{Remark 1}}:
\textit{FAS demonstrate a significant ability to mitigate HIs, achieving notable improvements in the minimum achievable rate. The deployment of FAS at both the BS and CUs maximizes performance and system robustness, while even equipping only the BS with FAS yields substantial gains under low transmit power conditions. These results confirm that FAS effectively alleviates the adverse impact of HIs and enhances multi-user system performance.}

In Fig.~\ref{RvsN}, we analyze the impact of the number of fluid antennas at the BS $N$ on the minimum achievable rate $R$. The results show that increasing $N$ consistently enhances the minimum achievable rate across all schemes, indicating that \textit{deploying more fluid antennas significantly improves system performance}. Besides, the rate of improvement is notably higher for the proposed scheme compared to others, demonstrating its superior efficiency in leveraging additional fluid antennas.

Notably, the inclusion of the FPA scheme with no HIs(\(\eta = 0\)) serves as an ideal baseline. The gap between this baseline and the other schemes highlights the detrimental impact of HIs. Our proposed scheme demonstrates robust performance gains in the presence of such impairments, showcasing an improved achievable rate that consistently outperforms existing methods. Moreover, the results emphasize that the proposed scheme achieves a balance between scalability and robustness, making it particularly effective in mitigating the adverse effects of HIs. These findings further demonstrate the potential of FAS to enhance multi-user communication system performance by exploiting the flexibility and adaptability offered by fluid antennas. This underscores \textbf{\textit{the capability of FAS to significantly alleviate the limitations imposed by HIs, especially as the number of fluid antennas increases.}}

In Fig.~\ref{RvsL}, we present the minimum achievable rate as a function of the number of paths $ L$  for various antenna schemes. As $L$  increases, all schemes demonstrate an improvement in the minimum achievable rate, with the proposed scheme consistently achieving the highest performance across all path numbers. This illustrates the effectiveness of the proposed approach in fully utilizing the available paths to enhance communication efficiency.

Fig.~\ref{RvsL} highlights that the proposed scheme outperforms the other schemes. The superior performance of the proposed scheme demonstrates its ability to better exploit multi-path diversity, which is crucial in complex wireless environments. This adaptability is particularly advantageous in wireless communication, where multi-path propagation can be leveraged to improve link robustness and reliability. By dynamically adjusting the antenna configurations, the proposed scheme maximizes the benefits of multi-path channels, resulting in a more robust and efficient communication system.

In Fig.~\ref{RvsA}, the impact of the moving region $A/\lambda$ on the minimum achievable rate is analyzed across different antenna schemes. As $A/\lambda$ increases, the minimum achievable rate improves for all schemes, with the proposed method consistently achieving the best performance. This result highlights the effectiveness of leveraging the expanded movement region to enhance communication efficiency.

A closer examination of the results for the EAS and TFA schemes reveals intriguing trends. Notably, when $A/\lambda$ is small (e.g., close to 1), the performance of the TFA scheme is comparable to or even wosre than the EAS scheme. \textit{This suggests that, under constrained movement regions, the inherent flexibility of fluid antennas may not be fully realized, as the limited movement reduces the spatial diversity benefits they provide.} Interestingly, as $A/\lambda$ increases, the performance of TFA surpasses EAS, emphasizing the advantage of fluid antennas when sufficient movement space is available. However, the proposed scheme, which equips both the BS and CUs with fluid antennas, consistently outperforms TFA and EAS, showcasing the compounded benefits of deploying fluid antennas at both ends.

These insights underscore a key characteristic of FAS: their ability to enhance system performance is closely tied to the availability of adequate movement space. In scenarios with a highly constrained moving region, fixed antenna selection strategies like EAS might even outperform FAS due to reduced electromagnetic effects and optimized static antenna placement. \textit{\textbf{Therefore, ensuring sufficient movement space is critical to unlocking the full potential of FAS in improving communication performance.}} Hence, we have the following remark:

\textbf{\emph{Remark 2}}:
\textit{The performance of FAS  is highly dependent on the availability of sufficient movement space. In constrained scenarios, fixed antenna selection methods like EAS may outperform FAS due to reduced electromagnetic effects. Thus, ensuring adequate movement space is essential for realizing the full potential of FAS.}

In Fig.~\ref{Rvseta}, we show the effect of the hardware degradation factor $\eta$ on the minimum achievable rate for different antenna schemes. As $\eta$ increases, indicating greater HIs, the minimum achievable rate decreases across all schemes. This trend highlights the negative impact of hardware degradation on communication performance.

The proposed scheme consistently achieves the highest minimum achievable rate, demonstrating its resilience to HIs compared to the TFA, EAS, RFA, and FPA schemes. Notably, while all schemes experience performance declines as $\eta$ grows, the proposed method maintains a distinct advantage, suggesting that it can better mitigate the effects of HIs. This robustness makes the proposed scheme more suitable for practical implementations where HIs are inevitable.

For instance, when $\eta = 0.05$, the proposed scheme achieves a minimum achievable rate of approximately $3.8$ bps/Hz, which is about $22.6\%$ higher than TFA (approximately $3.1$ bps/Hz) and over $80\%$ higher than the FPA scheme (approximately $2.1$ bps/Hz). Even at lower hardware degradation levels, such as \(\eta = 0.01\), the proposed scheme outperforms other schemes significantly, achieving a minimum rate of nearly $6.0$ bps/Hz, compared to around $5.2$ bps/Hz for TFA and $4.4$ bps/Hz for FPA. These observations provide strong evidence that FAS can make a measurable difference in mitigating the adverse effects of HIs, highlighting its effectiveness in preserving communication performance under practical conditions.

\section{Conclusion}\label{sec:conclusion}
In this paper, we have investigated the FAS-assisted multi-user communication system with transceiver HIs, where the BS with multiple FAs sends signals to multiple CUs each with a single FA. We proposed to maximize the minimum communication rate between all CUs under the constraints of the maximum transmit power of the BS, the finite moving region of FAs at the transceiver, and distance requirement between CUs. In particular, we proposed a BCD algorithm that integrates SCA, SRCR, and MM algorithms to alternately optimize the BS' beamforming, the BS' transmit FAs' positions, and each CU's receive FA's position. Simulation results confirmed the superiority of the FAS compared to FPAs, and also demonstrated the pragmatic importance of the transmit FA and receive FA in various scenarios.

\appendices
\section{Derivations of $\nabla {\tilde{\Upsilon}_k}(\mathbf{t}_n)$, $\nabla^2 {\tilde{\Upsilon}_k}(\mathbf{t}_n)$, and $\kappa_k^n$}\label{appendixA}
In this section, we provide the detail derivations of $\nabla {\tilde{\Upsilon}_k}(\mathbf{t}_n)$, $\nabla^2 {\tilde{\Upsilon}_k}(\mathbf{t}_n)$, and $\kappa_k^n$. For ${\tilde{\Upsilon}_k}(\mathbf{t}_n)$, we have
\begin{align}\label{a1}
{\tilde{\Upsilon}_k}(\mathbf{t}_n)=2\left( \sum_{l=1 }^{L_k^t}   \lvert   \bm{\xi}_k^{l}   \rvert  \cos\left(\varsigma_k^{l}(\mathbf{t}_n)\right)\right),
\end{align}
where
\begin{align}\label{a2}
\varsigma_k^{l}(\mathbf{t}_n)=\frac{2\pi}{\lambda}\varrho_{k,l}^t(\mathbf{t}_n)-\angle\bm{\xi}_k^{l}.
\end{align}

The gradient vector and the Hessian matrix of ${\tilde{\Upsilon}_k}(\mathbf{t}_n)$ can be respresented as $ \nabla {\tilde{\Upsilon}_k}(\mathbf{t}_n)=\left[\frac{\partial {\tilde{\Upsilon}_k}(\mathbf{t}_n)}{\partial x^{t}_n},\frac{\partial {\tilde{\Upsilon}_k}(\mathbf{t}_n)}{\partial y^{t}_n}  \right]$ and
$\nabla^2 {\tilde{\Upsilon}_k}(\mathbf{t}_n)=
\begin{bmatrix}
\frac{\partial {\tilde{\Upsilon}_k}(\mathbf{t}_n)}{\partial x^{t}_n \partial x^{t}_n} & \frac{\partial {\tilde{\Upsilon}_k}(\mathbf{t}_n)}{\partial x^{t}_n \partial y^{t}_n} \\
\frac{\partial {\tilde{\Upsilon}_k}(\mathbf{t}_n)}{\partial y^{t}_n \partial x^{t}_n} & \frac{\partial {\tilde{\Upsilon}_k}(\mathbf{t}_n)}{\partial y^{t}_n \partial y^{t}_n},
\end{bmatrix}$
, respectively. Details are presented as follows:
\begin{align}
\frac{\partial{\tilde{\Upsilon}_k}(\mathbf{t}_n)}{\partial x^{t}_n} =& - \frac{4\pi}{\lambda}\sum_{l=1}^{L_k^t} \lvert \bm{\xi}_k^{l} \rvert\sin\theta_{k,l}^t\cos\varphi_{k,l}^t\sin(\varsigma_k^{l}(\mathbf{t}_n)), \label{a3} \\
\frac{\partial{\tilde{\Upsilon}_k}(\mathbf{t}_n)}{\partial y^{t}_m} =& - \frac{4\pi}{\lambda}\sum_{l=1}^{L_k^t} \lvert \bm{\xi}_k^{l} \rvert \cos\phi_{k,;}^t\sin(\varsigma_k^{l}(\mathbf{t}_n)),  \label{a4} \\
\frac{\partial^2{\tilde{\Upsilon}_k}(\mathbf{t}_n)}{\partial x^{t}_n \partial x^{t}_n}=&\frac{8\pi^2}{\lambda^2}\left( \sum_{l=1}^{L_k^t}\lvert \bm{\xi}_k^{l} \rvert  \sin^2\theta_{k,l}^t \cos^2\varphi_{k,l}^t \cos(\varsigma_k^{l}(\mathbf{t}_n))    \right),  \label{a5} \\
\frac{\partial^2{\tilde{\Upsilon}_k}(\mathbf{t}_n)}{\partial x^{t}_n \partial y^{t}_n}=&
\frac{\partial^2{\tilde{\Upsilon}_k}(\mathbf{t}_n)}{\partial y^{t}_n \partial x^{t}_n}
=\frac{8\pi^2}{\lambda^2}\left( \sum_{l=1}^{L_k^t} \lvert \bm{\xi}_k^{l} \rvert  \sin\theta_{k,l}^t \cos\varphi_{k,l}^t \right. \nonumber\\ &\cdot \left.\cos\theta_{k,l}^t \cos(\varsigma_k^{l}(\mathbf{t}_n)) \right),  \label{a6} \\
\frac{\partial^2{\tilde{\Upsilon}_k}(\mathbf{t}_n)}{\partial y^{t}_n \partial y^{t}_n}=&\frac{8\pi^2}{\lambda^2}\left( \sum_{l=1}^{L_k^t} \lvert \bm{\xi}_k^{l} \rvert  \cos^2\theta_{k,l}^t \cos(\varsigma_k^{l}(\mathbf{t}_n)) \right).\label{a8}
\end{align}

After obtaining the Hessian matrix, we have
\begin{align} \label{a9}
\lvert \lvert\nabla^2 {\bm{\widetilde{\Upsilon}}_k}(\mathbf{t}_n)\rvert \rvert_2^2\leq &\lvert \lvert \nabla^2 {\bm{\widetilde{\Upsilon}}_k}(\mathbf{t}_n)\rvert \rvert_F^2 \nonumber \\
=&\left(\frac{\partial {\bm{\widetilde{\Upsilon}}_k}(\mathbf{t}_n)}{\partial x^{t}_n \partial x^{t}_n}\right)^2+\left(\frac{\partial {\bm{\widetilde{\Upsilon}}_k}(\mathbf{t}_n)}{\partial x^{t}_n \partial y^{t}_n}\right)^2 \nonumber \\
&+\left(\frac{\partial {\bm{\widetilde{\Upsilon}}_k}(\mathbf{t}_n)}{\partial y^{t}_n \partial x^{t}_n}\right)^2+\left(\frac{\partial {\bm{\widetilde{\Upsilon}}_k}(\mathbf{t}_n)}{\partial y^{t}_n \partial y^{t}_n}\right)^2.
\end{align}
Furthermore, due to $||\nabla^2 {\bm{\widetilde{\Upsilon}}_k}(\mathbf{t}_n)||_2\mathbf{I}_2 \succeq \nabla^2 {\bm{\widetilde{\Upsilon}}_k}(\mathbf{t}_n)$, we can choose $\kappa_k^n$ as
\begin{align}
\kappa_k^n=\frac{16\pi^2}{\lambda^2} \sum_{l=1}^{L_k^t} \lvert \bm{\xi}_k^{l} \rvert ,
\end{align}
which satisfies $\kappa_k^n \geq ||\nabla^2 {\bm{\widetilde{\Upsilon}}_k}(\mathbf{t}_n)||_2$, and $\kappa_k^n\mathbf{I}_2 \succeq \nabla^2 {\bm{\widetilde{\Upsilon}}_k}(\mathbf{t}_n)$.


\begin{thebibliography}{10}

\bibitem{CXWang23}
 C.-X. Wang {\em et al.}, ``On the road to 6G: Visions, requirements, key technologies and testbeds," \emph{IEEE Commun. Surveys Tuts.}, vol. 25, no. 2, pp. 905--974, Feb. 2023.

\bibitem{EGLarsson2014}
E. G. Larsson, O. Edfors, F. Tufvesson, and T. L. Marzetta, ``Massive MIMO for next generation wireless systems,'' \emph{IEEE Commun. Mag.}, vol.~52, no.~2, pp. 186--195, Feb. 2014.

\bibitem{ZWang2024}
Z. Wang {\em et al.}, ``Extremely large-scale MIMO: Fundamentals, challenges, solutions, and future directions,'' \emph{IEEE Wireless Commun.}, vol. 31, no. 3, pp. 117--124, Jun. 2024.

\bibitem{TWu2024}
T. Wu, {\em et al.}. "Fluid antenna systems enabling 6G: Principles, applications, and research directions." {\em arXiv preprint}, \url{arXiv:2412.03839}, 2024.

\bibitem{XLai24}
X. Lai, T. Wu, J. Yao, C. Pan, M. Elkashlan, and K.-K. Wong, ``On performance of fluid antenna system using maximum ratio combining," \emph{IEEE Commun. Lett.}, vol. 28, no. 2, pp. 402--406, 2024.

\bibitem{KKWong21}
K.-K. Wong, A. Shojaeifard, K.-F. Tong, and Y. Zhang, ``Fluid antenna system," \emph{IEEE Trans. Wireless Commun.}, vol. 20, no. 3, pp. 1950--1962, Mar. 2021.

\bibitem{JYao24}
J. Yao {\em et al.}, ``Proactive monitoring via jamming in fluid antenna systems,'' \emph{IEEE Commun. Lett.}, vol.~28, no.~7, pp. 1698--1702, Jul. 2024.

\bibitem{CWang24}
C. Wang {\em et al.}, ``Fluid antenna system liberating multiuser MIMO for ISAC via deep reinforcement learning,'' \emph{IEEE Trans. Wireless Commun.}, vol.~23, no.~9, pp. 10879--10894, Sep. 2024.

\bibitem{Yao20241} 
J. Yao {\em et al.}, ``A framework of FAS-RIS systems: Performance analysis and throughput optimization,'' {\em arXiv preprint}, \url{arXiv:2407.08141}, 2024. 

\bibitem{LZhu241}
L. Zhu, W. Ma, and R. Zhang, ``Modeling and performance analysis for movable antenna enabled wireless communications,'' \emph{IEEE Trans. Wireless Commun.}, vol.~23, no.~6, pp.~6234--6250, Jun. 2024.

\bibitem{WMa241}
W. Ma, L. Zhu, and R. Zhang, ``MIMO capacity characterization for movable antenna systems,'' \emph{IEEE Trans. Wireless Commun.}, vol.~23, no.~4, pp. 3392--3407, Apr. 2024.

\bibitem{YGao24}
Y. Gao, Q. Wu, and W. Chen, ``Joint transmitter and receiver design for movable antenna enhanced multicast communications,'' \emph{IEEE Trans. Wireless Commun.}, early access, \url{doi:10.1109/TWC.2024.3463390}, 2024.

\bibitem{LZhu20242}
L. Zhu, W. Ma, Z. Xiao, and R. Zhang, ``Performance Analysis and Optimization for Movable Antenna Aided Wideband Communications,'' \emph{IEEE Trans. Wireless Commun.}, vol. 23, no. 12, pp. 18653--18668, Dec. 2024.

\bibitem{GHu2024}
G. Hu, Q. Wu, K. Xu, J. Si, and N. Al-Dhahir, ``Secure wireless communication via movable-antenna array,'' \emph{IEEE Signal Process. Lett.}, vol.~31, pp. 516--520, Jan. 2024.

\bibitem{WMei2024}
W. Mei, X. Wei, B. Ning, Z. Chen, and R. Zhang, ``Movable-antenna position optimization: A graph-based approach,'' \emph{IEEE Wireless Commun. Lett.}, vol.~13, no.~7, pp. 1853--1857, Jul. 2024.

\bibitem{SShen2017}
S.~Shen, Y.~Sun, S.~Song, D.~P. Palomar, and R.~D. Murch, ``Successive Boolean optimization of planar pixel antennas,'' \emph{IEEE Trans. Antennas \& Propag.}, vol.~65, no.~2, pp. 920--925, Feb. 2017.
\bibitem{FJiang2022}
F.~Jiang {\em et al.}, ``Pixel antenna optimization based on perturbation sensitivity analysis,'' \emph{IEEE Trans. Antennas \& Propag.}, vol.~70, no.~1, pp. 472--486, Jan. 2022.

\bibitem{KKWong2022}
K.-K. Wong {\em et al.}, ``Closed-form expressions for spatial correlation parameters for performance analysis of fluid antenna systems,'' {\em IET Elect. Lett.}, vol. 58, no. 11, pp. 454--457, Apr. 2022.

\bibitem{MKhammassi23}
M. Khammassi, A. Kammoun, and M.-S. Alouini, ``A new analytical approximation of the fluid antenna system channel,'' \emph{IEEE Trans. Wireless Commun.}, vol. 22, no. 12, pp. 8843--8858, Dec. 2023.

\bibitem{PR24}
P. Ramirez-Espinosa, D. Morales-Jimenez, and K.-K. Wong, ``A new spatial block-correlation model for fluid antenna systems,'' {\em IEEE Trans. Wireless Commun.}, vol. 23, no. 11, pp. 15829--15843, Nov. 2024.

\bibitem{CSkouroumounis2023}
C. Skouroumounis and I. Krikidis, ``Fluid antenna with linear MMSE channel estimation for large-scale cellular networks,'' {\em IEEE Trans. Commun.}, vol. 71, no. 2, pp. 1112--1125, Feb. 2023.

\bibitem{WKNew20241}
W. K. New, K.-K. Wong, H. Xu, K.-F. Tong, and C.-B. Chae, ``Fluid antenna system: New insights on outage probability and diversity gain,'' \emph{IEEE Trans. Wireless Commun.}, vol. 23, no. 1, pp. 128--140, Jan. 2024.

\bibitem{JD2024}
J. D. Vega-Sánchez  {\em et al.}, ``Fluid antenna system: Secrecy outage probability analysis,'' {\em IEEE Trans. Veh. Technol.}, vol. 73, no. 8, pp. 11458--11469, Aug. 2024.

\bibitem{HXu20242}
H. Xu  {\em et al.}, ``Revisiting outage probability analysis for two-user fluid antenna multiple access system,'' \emph{IEEE Trans. Wireless Commun.}, vol.~23, no.~8, pp.~9534--9548, Aug. 2024.

\bibitem{LTlebaldiyeva2023}
L. Tlebaldiyeva, S. Arzykulov, T. A. Tsiftsis, and G. Nauryzbayev, ``Full-duplex cooperative NOMA-based mmWave networks with fluid antenna system (FAS) receivers,'' in {\em Proc. Int. Balkan Conf. Commun. Netw. (BalkanCom)}, Istanbul, Turkey, Jun. 2023, pp. 1--6.

\bibitem{JZheng2024}
J. Zheng {\em et al.}, ``FAS-assisted NOMA short-packet communication systems,'' {\em  IEEE Trans. Veh. Technol.}, vol.~73, no.~7, pp. 10732--10737, Jul. 2024.

\bibitem{New2024}
W.-K. New  {\em et al.}, ``Fluid antenna system enhancing orthogonal and nonorthogonal multiple access,'' \emph{IEEE Commun. Lett.}, vol.~28, no.~1, pp. 218--222, Jan. 2024.

\bibitem{JTang2024}
J. Tang, C. Pan, Y. Zhang, H. Ren, and K. Wang, ``Secure MIMO communication relying on movable antennas,'' \emph{IEEE Trans. Commun.}, early access, \url{doi:10.1109/TCOMM.2024.3465369}, 2024.

\bibitem{YYe2024}
Y. Ye, L. You, J. Wang, H. Xu, K.-K. Wong, and X. Gao, “Fluid antenna-assisted MIMO transmission exploiting statistical CSI,” \emph{IEEE Commun.
Lett.}, vol. 28, no. 1, pp. 223--227, Jan. 2024.

\bibitem{ZCheng2024}
Z. Cheng, N. Li, J. Zhu, X. She, C. Ouyang, and P. Chen, ``Sum-rate maximization for fluid antenna enabled multiuser communications,'' \emph{IEEE Commun. Lett.}, vol.~28, no.~5, pp. 1206--1210, May 2024.

\bibitem{GHu20242}
G. Hu {\em et al.}, ``Fluid antennas-enabled multiuser uplink: A low-complexity gradient descent for total transmit power minimization,'' \emph{IEEE Commun. Lett.},  vol. 28, no. 3, pp. 602--606, Mar. 2024.

\bibitem{NLi2024}
N. Li, W. Mei, B. Ning, and P. Wu, ``Movable antenna enhanced AF relaying: Two-stage antenna position optimization'', {\em arXiv preprint} \url{arXiv:2408.05746}, 2024.

\bibitem{JYao20242}
J. Yao, L. Zhou, T. Wu, M. Jin, C. Huang, and C. Yuen, ``FAS vs. ARIS: Which is more important for FAS-ARIS communication systems?'', {\em arXiv preprint} \url{arXiv:2408.09067}, 2024.


\bibitem{EB2014}
E. Bjärnson, J. Hoydis, M. Kountouris, and M. Debbah, ``Massive MIMO systems with non-ideal hardware: Energy efficiency, estimation, and capacity limits,'' \emph{IEEE Trans. Inf. Theory}, vol. 60, no. 11, pp. 7112--7139, Nov. 2014.

\bibitem{ZLiu2020}
Z. Liu, G. Lu, Y. Ye, and X. Chu, ``System outage probability of PS-SWIPT enabled two-way AF relaying with hardware impairments,'' \emph{IEEE Trans. Veh. Technol.}, vol. 69, no. 11, pp. 13532--13545, Nov. 2020.

\bibitem{ZXing2021}
Z. Xing, R. Wang, J. Wu, and E. Liu, ``Achievable rate analysis and phase shift optimization on intelligent reflecting surface with hardware impairments,'' \emph{IEEE Trans. Wireless Commun.}, vol. 20, no. 9, pp. 5514--5530, Sep. 2021.

\bibitem{GZhou2021}
G. Zhou, C. Pan, H. Ren, K. Wang, and Z. Peng, ``Secure wireless communication in RIS-aided MISO system with hardware impairments,'' \emph{IEEE Wireless Commun. Lett.}, vol. 10, no. 6, pp. 1309--1313, Jun. 2021.

\bibitem{AP2021}
A. Papazafeiropoulos, E. Björnson, P. Kourtessis, S. Chatzinotas, and J. M. Senior, “Scalable cell-free massive MIMO systems: Impact of hardware impairments,” \emph{IEEE Trans. Veh. Technol.}, vol. 70, no. 10, pp. 9701--9715, Oct. 2021.

\bibitem{JWang2023}
J. Wang, S. Gong, Q. Wu, and S. Ma, ``RIS-aided MIMO systems with hardware impairments: Robust beamforming design and analysis,'' {\em IEEE Trans.  Wireless Commun.}, vol. 22, no. 10, pp. 6914--6929, Oct. 2023.

\bibitem{JFang2023}
J. Fang, C. Zhang, Q. Wu, and A. Li, ``Improper Gaussian signaling for IRS assisted multiuser SWIPT systems with hardware impairments,'' {\em  IEEE Trans. Veh. Technol.}, vol. 72, no. 10, pp. 13024--13038, Oct. 2023.

\bibitem{JDai2024}
J. Dai, J. Ye, K. Wang, C. Pan, and H. Fan, ``Joint radar-communication beamforming considering both transceiver hardware impairments  and imperfect CSI," \emph{IEEE Wireless Commun. Lett.}, vol.~13, no.~7, pp. 1898--1902, Jul. 2024.

\bibitem{QLi2024}
Q. Li, M. El-Hajjar, Y. Sun, and L. Hanzo, ``Performance analysis of reconfigurable holographic surfaces in the near-field scenario of cell-free networks under hardware impairments,'' {\em IEEE Trans. Wireless Commun.}, vol. 23, no. 9, pp. 11972--11984, Sep. 2024.

\bibitem{ZPeng2024}
Z. Peng, Z. Zhang, C. Pan, M. D. Renzo, O. A. Dobre, and J. Wang, ``Beamforming optimization for active RIS-aided multiuser communications with hardware impairments,'' {\em IEEE Trans. Wireless Commun.}, vol. 23, no. 8, pp. 9884--9898, Aug. 2024.

\bibitem{HLi2024}
H. Li, Y. Ye, L. Lv, G. Lu, and N. Al-Dhahir, ``Covert cooperative backscatter communications with hardware impairments,'' {\em  IEEE Trans. Veh. Technol.}, vol. 73, no. 7, pp. 10150--10163, Jul. 2024.

\bibitem{MGrant}
M. Grant and S. Boyd, ``CVX: MATLAB software for disciplined convex programming,'' [Online]. Available: \url{http://cvxr.com/cvx}.

\bibitem{LZhou24}
L. Zhou {\em et al.}, ``Fluid antenna-assisted ISAC systems,'' {\em IEEE Wireless Commun. Lett.}, vol. 13, no. 12, pp. 3533--3537, Dec. 2024.

\bibitem{JYao123}
J. Yao {\em et al.}, "FAS-driven spectrum sensing for cognitive radio networks," {\em IEEE Internet Things J.}, early access, \url{doi: 10.1109/JIOT.2024.3518623}, 2024.

\bibitem{SBoyd}
S. Boyd and L. Vandenberghe, {\em Convex Optimization}. Cambridge, U.K.: Cambridge Univ. Press, 2004.

\bibitem{TWu20241}
T. Wu {\em et al.}, ``Joint angle estimation error analysis and 3-D positioning algorithm design for mmWave positioning system,'' {\em IEEE Internet
Things J.}, vol. 11, no. 2, pp. 2181--2197, Jan. 2024.

\end{thebibliography}
\end{document}